%% file: ADM_paper.tex
\documentclass[11pt, a4paper]{article}
\usepackage{jheppub}
\usepackage{subcaption}
\usepackage{multirow}
\usepackage{multicol}
\usepackage[normalem]{ulem}
\usepackage{float}
\usepackage{caption}
\usepackage[english]{babel}
\usepackage{adjustbox}
\usepackage{tikz-feynman}
\usepackage{contour}
\usepackage[toc,page]{appendix}
\usepackage{url}
\usepackage{slashed}

\newcommand{\br}{\begin{eqnarray}}
\newcommand{\er}{\end{eqnarray}}

\def \MET{\slashed{E}_T}

\usepackage{tikz,xcolor,hyperref}

\definecolor{lime}{HTML}{A6CE39}
\DeclareRobustCommand{\orcidicon}{\hspace{-4pt}
\begin{tikzpicture}
	\draw[lime, fill=lime] (0,0) 
	circle [radius=0.16] 
	node[white] {\hspace{0.1mm}{\fontfamily{qag}\selectfont \tiny ID}};
	\draw[white, fill=white] (-0.07,0.1) 
	circle [radius=0.01];
\end{tikzpicture}
\hspace{-3.2mm}
}

\foreach \x in {A, ..., Z}{\expandafter\xdef\csname 
orcid\x\endcsname{\noexpand\href{https://orcid.org/\csname orcidauthor\x\endcsname}
	{\noexpand\orcidicon}}
}

\title{\boldmath Constraining Asymmetric Dark Matter using Colliders and Direct Detection}

\author[a,b]{Arnab Roy\orcidA{},\,}
\author[a]{Basudeb Dasgupta\orcidB{},\,}
\author[a]{and Monoranjan Guchait\orcidC{}}
\affiliation[a]{\small
	Tata Institute of Fundamental Research, Homi Bhabha Road, Colaba, Mumbai 400005, India}
\affiliation[b]{School of Physics and Astronomy, Monash University,\\
	Wellington Road, Clayton, Victoria 3800, Australia}

\emailAdd{arnab.roy1@monash.edu}
\emailAdd{bdasgupta@theory.tifr.res.in}
\emailAdd{guchait@tifr.res.in}

\abstract{We reappraise the viability of asymmetric dark matter (ADM) realized as a Dirac fermion coupling dominantly to the Standard Model fermions. Treating the interactions of such a DM particle with quarks/leptons in an effective-interactions framework, we derive updated constraints using mono-jet searches from the Large Hadron Collider (LHC) and mono-photon searches at the Large Electron-Positron (LEP) collider. We carefully model the detectors used in these experiments, which is found to have significant impact. The constraint of efficient annihilation of the symmetric part of the ADM, as well as other observational constraints are synthesized to produce a global picture. Consistent with previous work, we find that ADM with mass in the range~~\mbox{1--100\,GeV} is strongly constrained, thus ruling out its best motivated mass range. However, we find that leptophilic ADM remains allowed for $\gtrsim 10$\,GeV DM, including bounds from colliders, direct detection, and stellar heating. We forecast that the Future Circular Collider for electron-positron collisions (FCC-ee) will improve sensitivity to DM-lepton interactions by almost an order of magnitude.}

\begin{document}
\maketitle
\flushbottom
	
\section{Introduction}
Despite the significant asymmetry between the cosmological abundances of matter and antimatter, we observe comparable abundances of ordinary matter, dominated by baryons\,(b), and of dark matter (DM), namely $\Omega_{\rm b}\simeq \Omega_{\rm DM}/5$~\cite{Aghanim:2018eyx}. This observation has motivated the Asymmetric Dark Matter (ADM) hypothesis~\cite{Nussinov:1985xr,Gelmini:1986zz,Zurek:2013wia,Petraki:2013wwa}. Here, the similar baryon and DM abundances are a consequence of both densities arising from an asymmetry of the corresponding Dirac fermion over its anti-fermion, with the two asymmetries related to each other in some way~\mbox{\cite{Barr:1990ca,Barr:1991qn,Kaplan:1991ah,Gudnason:2006ug,Gudnason:2006yj,Kribs:2009fy}}. Similar to models of weakly interacting massive particle type DM, where the DM is produced via thermal freeze-out and traditionally though of as having mass near the electroweak scale, the models for ADM are also testable at particle colliders because the expected mass-range for ADM particles is around the GeV scale or close to it. This is because ADM models predict that the number abundances, $n_{\rm DM}$ and $n_{\rm b}$, are similar  and the ratio of observed mass abundances then predicts ADM particles at the GeV-scale~\cite{Kaplan:2009ag,Cohen:2009fz,Cohen:2010kn,An:2009vq,Chun:2010hz,Shelton:2010ta,Davoudiasl:2010am,Haba:2010bm,Gu:2010ft,Blennow:2010qp}. 

The ADM hypothesis therefore presents a theoretical alternative to WIMPs, retaining almost all of its strengths. These models are well-motivated from the ultraviolet considerations, such as GUTs and additional confined gauge sectors~\cite{Agashe:2004bm,Kitano:2005ge,Cosme:2005sb,Suematsu:2005kp,Tytgat:2006wy,Banks:2006xr,Page:2007sh,Hooper:2004dc,Kitano:2004sv,Belyaev:2010kp}. They are conceptually minimal, employing the mathematical structures and physical mechanisms that are believed to already underlie the Standard Model, e.g., Dirac fermions, particle-antiparticle asymmetry, gauge interactions, anomalies, and sphalerons. And importantly, they predict GeV-scale masses for the DM particle, with observable signals not only at direct detection~(DD) experiments and colliders, but also constraints from astrophysics and cosmology.

While considerable attention has been devoted to the theories of ADM, its detectability in particle colliders remains relatively less well-explored. Two previous studies, by Buckley~\cite{Buckley:2011kk} and by March-Russell et al.\,(2012)~\cite{March-Russell:2012elz}, presented collider and DD constraints on ADM, together with the constraints imposed by the model itself. The key idea being that efficient annihilation of the symmetric part of the ADM relic density requires that the coupling between DM and SM fermions cannot be too weak. This is under the usual simplifying assumption that the dark matter annihilates into and scatters off SM fermions, and these processes are mediated by the same elementary interaction. Quite remarkably, already in this early work, it was found that collider data and direct detection searches practically rule out ADM with mass all the way up to 100 GeV or more.

Developments over the past decade motivate a reappraisal. Continued data-taking at the LHC as well as at the significantly more sensitive and diverse set of DD searches now allow a more comprehensive study of ADM couplings to SM fermions. Further, there is a renewed interest in ADM, especially in connection to their dramatic impact on compact stars~\cite{McDermott:2011jp,Kouvaris:2011fi,Kouvaris:2011gb,Kouvaris:2012dz,Bramante:2013hn,Bell:2013xk,Goldman:2013qla,Bramante:2013nma,Bramante:2014zca,Bramante:2015dfa,Kouvaris:2015rea,Bramante:2017ulk,Kouvaris:2018wnh,Gresham:2018rqo,Garani:2018kkd,Dasgupta:2020dik,Dasgupta:2020mqg,Rutherford:2022xeb,Bhattacharya:2023stq,Bramante:2023djs}. And finally, several new DM experiments~\cite{Cebrian:2022brv}, as well as experiments at the HL-LHC, aim to search widely for DM, which motivate an updated idea of \emph{`where to look for new physics'.} 

In this study, we revisit collider constraints on ADM and provide new and updated bounds on the interactions between ADM and quarks/leptons. Our analysis includes a careful incorporation of detector efficiencies, corresponding to different phase space regions of the final state particles.  {For DM-quark interactions}, our collider limits are based on the outcomes of mono-jet searches conducted at the LHC. Further, by combining other DD and collider bounds on DM with the prerequisites of the ADM hypothesis, we find that all the operators we consider are ruled out for DM mass up to a few hundred GeV; for some operators the exclusion stretches almost to TeV masses. Prima facie the basic ADM scenario appears strongly disfavored.

We then turn our attention to a leptophilic variant of the ADM scenario, where ADM couples to SM leptons but not to quarks. Such ADM models may be motivated by theory,  and are poised to take advantage of the upcoming experiments.  In some ADM models, the DM asymmetry fundamentally arises due to new sphalerons that transform the lepton asymmetry created by leptogenesis into a DM asymmetry. Here, a chiral gauge interaction shared by DM and leptons directly leads to the above phenomenology and does not necessarily need to involve quarks. On the observational side, DM interactions with leptons (electrons, in particular) has gained interest over the past decade. Unsurprisingly, DM-lepton interactions have been studied in depth ~\cite{Baltz:2002we,Chen:2008dh,Fox:2008kb,Cao:2009yy,Cohen:2009fz,Ibarra:2009bm,Bi:2009uj,Kopp:2009et,Ko:2010at,Chao:2010mp,Schmidt:2012yg,Das:2013jca,Schwaller:2013hqa,Bai:2014osa,Agrawal:2014ufa,Kopp:2014tsa} and a gamut of experiments have already produced interesting limits on leptophilic DM~\cite{XENON100:2015tol, LZ:2021xov, Chen:2018vkr, Marsicano:2018vin}. More importantly, there are now several novel proposals that promise to make significant progress in this direction in the near future~\cite{CRESST:2017ues,SuperCDMS:2020aus,SENSEI:2020dpa,DAMIC-M:2023gxo,aguilar2022oscura,Marshall:2020azl,Hertel:2018aal,SPICE:2023aqd,2020JLTP,Hochberg:2015pha,Hochberg:2019cyy,Geilhufe:2019ndy,Kahn:2021ttr,Blanco:2022cel,Huang:2023nms,Das:2023cbv,Carew:2023qrj}. Limits on DM-lepton interactions from the heating of neutron stars (NS)~\cite{Bell:2019pyc} and white dwarfs (WD)~\cite{Bell:2021fye} happen to provide some of the strongest constraints. On the collider front, mono-photon searches at LEP have been set bounds on four effective operators (three of which are independent)~\cite{Fox:2011fx}, which are more stringent than the NS or WD bounds for DM masses in the $\lesssim 100$ GeV range. Investigations have also focused on the prospects of identifying leptophilic dark matter through future lepton colliders, mainly the International Linear Collider (ILC)~\cite{Dutta:2017ljq,Barman:2021hhg,Kundu:2021cmo}. In the latter part of this paper, we update the bounds on {DM-lepton interactions} using the LEP mono-photon searches, compare with other constraints as for the quark case, and appraise the viability of leptophilic ADM. We also forecast discovery prospects at the prospective FCC-ee.

The structure of this paper is as follows. In Section 2, we start by establishing the EFT framework and provide a brief introduction to the ADM hypothesis and review the existing constraints on interactions between ADM and quarks. In Section 3, we calculate new bounds on the ADM-quark interactions from the mono-jet searches at the LHC and present the constraints on ADM-quark interactions for all EFT operators. Additionally, in Section 4, we investigate the DM-lepton interactions. Here, a comprehensive methodology to calculate the LEP bounds are presented, followed by the FCC-ee discovery prospects. The combined pictures of the allowed (and excluded) regions of ADM-lepton interactions from all relevant searches are presented at the end of section 4. Lastly, Section 5 summarizes our findings and concludes.

	\section{Dark Matter Interactions and Production}
	
\subsection{Effective interactions}	
	We assume the DM is a Dirac fermion and the mediators responsible for its interactions with the SM particles are heavier than the DM itself. The interactions are thus encoded in higher-dimensional operators within the framework of an effective field theory (EFT) involving SM and DM degrees of freedom~\cite{Beltran:2010ww,Goodman:2010ku}. However, we will ignore the dimension-5 couplings of DM to the SM Higgs, which is already very well studied and constrained. See, for example, the stringent upper limits of $\rm H_{SM}\to invisible$ branching fractions \cite{CMS:2022qva,ATLAS:2022yvh}.\footnote{Note that the Higgs portal for fermionic DM of mass smaller than $\sim$20 GeV, consistent with $\rm BR(H_{SM}\to invisible)<10\%$, leads to DM-nucleon scattering cross-section below the ``neutrino floor''~\cite{Strigari:2009bq,Billard:2013qya,Monroe:2007xp}, which is uninteresting from DD perspective~\cite{Arcadi:2021mag}. Alternatively, though a higher mass region of DM allows parameter regions above the neutrino floor, they are excluded by the DD limits~\cite{Arcadi:2021mag}.} With these considerations, in Table~\ref{tab:operators} we write down all allowed EFT operators at dimension-6. These operators are of the form
    \br
        \mathcal{O}_{\Gamma\Gamma^{'}} = \frac{1}{\Lambda^2}(\bar{\psi}\Gamma\psi)(\bar{\chi}\Gamma'\chi),
    \er
where $\psi$ and $\chi$ are Dirac fermion SM and DM particles, respectively, whereas $\Gamma^{(')}$s are strings of gamma matrices,
	\br
	\Gamma=\{1,\gamma^5,\gamma^\mu,\gamma^\mu\gamma^5,\sigma^{\mu\nu},\sigma^{\mu\nu}\gamma^5\}.
	\er
Among the tensor operators, $\bar{\psi}\sigma^{\mu\nu}\psi\bar{f}i\sigma_{\mu\nu}\gamma^5f$ is rewritable as $\bar{\psi}i\sigma^{\mu\nu}\gamma^5\psi\bar{f}\sigma_{\mu\nu}f\equiv\mathcal{O}_{pt}$ and $\bar{\psi}i\sigma^{\mu\nu}\gamma^5\psi\bar{f}i\sigma_{\mu\nu}\gamma^5f$ as $\bar{\psi}\sigma^{\mu\nu}\psi\bar{f}\sigma_{\mu\nu}f\equiv\mathcal{O}_{tt}$. So they are not separately considered. Moreover, operators enter the Lagrangian along with respective Wilson coefficients (WC). In this work, we will consider that the WCs are absorbed into the $\Lambda$. It is to be noted that while $\Lambda$ generally sets the upper limit for momentum transfer where the operator expansion in the EFT remains reliable, it does not establish a strict cut-off. A precise determination of EFT cut-off requires detailed knowledge of the masses and interaction strengths within the UV theory~\cite{Busoni:2013lha,Contino:2016jqw}. Nevertheless, a rough estimation of EFT validity is always feasible. For instance, assuming maximal coupling in the UV theory, one can derive the condition for validity of the effective interactions. For example, in our case it turns out to be $\Lambda\gtrsim m_{\chi}/2\pi$ with an assumption of s-channel momentum transfer~\cite{Busoni:2013lha,Shoemaker:2011vi}.
		
	\begin{table}
		\caption{\small List of dimension-6 operators connecting SM fermions to Dirac DM with flavor-universal couplings, along with the dependence of the scattering cross section on the spin of the SM target (and the leading dependence on $q$ or $v$), as well as the leading partial-wave contribution to the corresponding annihilation rate; $q$ and $v$ denote the magnitude of transferred momentum and the relative velocity between the target and the DM, respectively. }
		\begin{center}
			\begin{tabular}{ c c c c}
				\hline
				\hline
				Operator & Definition & Scattering  & Annihilation \\ 
				\hline
				\hline 
				$\mathcal{O}_{ss}$ & $\bar{\psi}\psi\bar{f}f$ & SI ~~ $(1)$ & $p$-wave\\  
				$\mathcal{O}_{pp}$ & $\bar{\psi}\gamma^5\psi\bar{f}\gamma^5f$ &  SD ~~$(q^2)$ &$s$-wave\\
				$\mathcal{O}_{sp}$ & $\bar{\psi}\psi\bar{f}i\gamma^5f$ & SD  ~~$(q)$ &$p$-wave\\ 
				$\mathcal{O}_{ps}$ & $\bar{\psi}i\gamma^5\psi\bar{f}f$ & SI  ~~$(q)$ &$s$-wave\\ 
				$\mathcal{O}_{vv}$ & $\bar{\psi}\gamma^{\mu}\psi\bar{f}\gamma_{\mu}f$ & SI  ~~ $(1)$ &$s$-wave\\
				$\mathcal{O}_{aa}$ & $\bar{\psi}\gamma^{\mu}\gamma^5\psi\bar{f}\gamma_{\mu}\gamma^5f$ &  SD ~~$(1)$ &$s$-wave $\propto m_q^2/m_{\chi}^2$\\
				$\mathcal{O}_{va}$ &  $\bar{\psi}\gamma^{\mu}\psi\bar{f}\gamma_{\mu}\gamma^5f$ &  SD ~~ $(v)$ &$s$-wave\\
				$\mathcal{O}_{av}$ & $\bar{\psi}\gamma^{\mu}\gamma^5\psi\bar{f}\gamma_{\mu}f$ &  SD ~~$(v)$ &$p$-wave\\
				$\mathcal{O}_{tt}$ & $\bar{\psi}\sigma^{\mu\nu}\psi\bar{f}\sigma_{\mu\nu}f$ & SD  ~~$(1)$ &$s$-wave\\
				$\mathcal{O}_{pt}$ & $\bar{\psi}i\sigma^{\mu\nu}\gamma^5\psi\bar{f}\sigma_{\mu\nu}f$ &  SI ~~$(q)$ &$s$-wave\\
				\hline
				\hline
			\end{tabular}
		\end{center}
		\label{tab:operators}
	\end{table}

Using available data, we will determine the viability of each of these operators as a function of the DM mass $m_\chi$ and the EFT expansion scale $\Lambda$. Note that the operators, $\mathcal{O}_{ss}$, $\mathcal{O}_{vv}$, $\mathcal{O}_{aa}$ and $\mathcal{O}_{tt}$ present velocity-unsuppressed interactions with nucleon, while the other interactions are suppressed. On the other hand, operators marked as SI are spin independent (i.e., do not depend on the spin of the target SM fermion), and thus allow more sensitive searches that can take advantage of coherent enhancement. For a detailed discussion of these non-relativistic dependencies of DM, see Ref.~\cite{Fan:2010gt}. The annihilation rate, $\langle \sigma v\rangle$, also depends on the velocity of DM. While \textit{s}-wave annihilation is independent of velocity, \textit{p}-wave annihilation is velocity-dependent and thus suppressed. As a result, $\mathcal{O}_{ss}$, $\mathcal{O}_{sp}$, and $\mathcal{O}_{av}$ lead to a velocity-suppressed annihilation rate. It is worth noting that, although annihilation for $\mathcal{O}_{aa}$  occurs through $s$-wave, it is suppressed by a factor proportional to $m_q^2/m_{\chi}^2$.

\subsection{Relic density and the asymmetry criterion}
	
The asymmetric dark matter paradigm assumes that both DM ($\chi$) and the anti-DM ($\bar{\chi}$) are present up to some era of the evolution of the Universe, as in the case of baryonic matter. Then by some mechanism, an asymmetry is generated between DM and anti-DM sectors, which gets frozen out with the departure from thermal equilibrium, and the symmetric part gets annihilated out. In this study we remain agnostic about the asymmetry generation and sharing mechanisms.
	
The total dark-abundance can be written as the sum of the yields of DM and anti-DM, i.e., $Y_{\chi}+Y_{\overline{\chi}}$, where the yield $Y=n/s$ is defined as the number density scaled by the entropy density of the early universe. This, in turn can be expressed as a symmetric and an asymmetric component~\cite{Graesser:2011wi,Iminniyaz:2011yp}:
	\br 
	Y_{\rm tot}=Y_{\chi}+Y_{\overline{\chi}}=(Y_{\chi}-Y_{\overline{\chi}})+2Y_{\overline{\chi}}=	Y_{\rm asy}+Y_{\rm sym},
	\er
where, $Y_{\rm asy}$ is the asymmetric yield, and $Y_{\rm sym}$ is the yield of the symmetric component.

For the symmetric part, following \cite{Iminniyaz:2011yp}, we take
	\br 
	Y_{\rm sym}=2Y_{\overline{\chi}}=\frac{2Y_{\rm asy}}{\exp\left[Y_{\rm asy}\lambda\left(\frac{a}{x_{F}}+\frac{3b}{x_F^2}\right)\right]-1},
	\label{eq:ydmbar}
	\er
	where $a$ and $b$ are coefficients in the partial wave expansion of $\langle \sigma v\rangle$, $\lambda$ is a function of DM mass:
	\br
	\lambda=\frac{4\pi}{\sqrt{90}}m_{\chi}M_{\rm pl}\sqrt{g_*},
	\er
	with reduced Planck mass $M_{\rm pl}=2.4\times 10^{18}$ GeV, and $g_*\sim 100$ being the number of available relativistic degrees of freedom at freeze-out. Also, the shifted coordinate $x_F$ is given by~\cite{Iminniyaz:2011yp}
	\br 
	x_{\rm F}=x_{\rm F_{0}}\left(1+0.285\frac{a\lambda Y_{\rm asy}}{x_{\rm F_{0}}^3}+1.35\frac{b\lambda Y_{\rm asy}}{x_{\rm F_{0}}^4}\right),
	\label{eq:xf}
	\er
with $x_{\rm F_{0}}=m_\chi/T_{\rm F}$ being the usual coordinate at freeze-out. The difference between $x_{\rm F}$ and $x_{\rm F_{0}}$ encodes the ADM correction to the $\bar\chi$ abundance, which is otherwise under-predicted by the standard WIMP treatment. The expressions of the annihilation cross-sections times velocity for each of the operators in Table \ref{tab:operators} are given in Appendix A. 
	
The constraint imposed in ADM model is that the symmetric part, as in Eq.\,\ref{eq:ydmbar},  must be decidedly sub-leading to the asymmetric part. We implement this by restricting $Y_{\rm sym}$ to contribute $\leq 1\%$ (say) to $\Omega_{\rm DM} h^2$ \cite{March-Russell:2012elz}:
	\br 
	Y_{\rm sym}\leq \frac{1}{100}\times \frac{\Omega_{\rm DM} h^2}{2.76\times 10^8}\left(\frac{\rm GeV}{m_{\chi}}\right).
	\label{eq:constraint}
	\er
This is to be achieved by annihilating out the symmetric component ($Y_{\rm sym}\to 0$) into SM fermions using the operators presented in Table~\ref{tab:operators}. When there is a stronger interaction between DM and SM particles, it becomes easier to annihilate the symmetric part. However, it also makes more challenging to pass the constraints set by various experiments. For instance, at a specific value of the DM mass $m_{\chi}$, a large $\Lambda$ implies less effective scattering interactions (SI or SD) with quarks, but at the same time, it also indicates less efficient annihilation in the early universe. Consequently, the requirement for efficient annihilation sets an upper limit on $\Lambda$ (say, $\Lambda<\Lambda_{1}$) via Eq.\,\ref{eq:constraint}. In contrast, null results from DD experiments establish a lower limit ($\Lambda>\Lambda_{2}$). The space that lies in-between, if it exists (i.e., $\Lambda_1>\Lambda_2$), defines the permissible range for ADM scenarios. However, if one finds $\Lambda_1\leq\Lambda_2$ in a particular mass range, it rules out the feasibility of an ADM scenario within that region. Operators that lead to $p$-wave annihilation, such as $\mathcal{O}_{ss}$, $\mathcal{O}_{sp}$, and $\mathcal{O}_{av}$, require a lower value of $\Lambda$ to satisfy the ADM requirement, resulting in stronger exclusions.

	\section{Constraining ADM-Quark Interactions}
	In this section, we first revisit the constraints on the DM-quark effective interactions and then find their implications on ADM scenario. The DM-quark interactions are mainly constrained by the DD experiments and the results of mono-jet searches at the LHC, where direct interaction of DM with quarks is present. These constraints were already shown in Ref.~\cite{Buckley:2011kk,March-Russell:2012elz}. More extensive experimental results and several new searches have come up since, which are expected to impose stronger and more robust constraints. In this section, we update the bounds derived from mono-jet searches at LHC using latest data and by taking relevant detector effects into account. We then present the big picture for ADM-quark interactions, comparing with constraints from the asymmetry, and including other important constraints, especially those from recent DD experiments.
	
\subsection{Mono-jet searches at hadron colliders}
\begin{figure}
		\centering
		\begin{tikzpicture}[scale=1.3]
		\begin{feynman}
		\vertex[blob] (m) at ( 0, 0) {\contour{white}{}};
		\vertex (a) at (-2,-1.5) {$q$};
		\vertex (b) at ( 2,-1.5) {$\bar{\chi}$};
		\vertex (c) at (-2, 1.5) {$\bar{q}$};
		\vertex (d) at ( 2, 1.5) {$\chi$};
		\vertex (e) at (-0.8, 0.44) {};
		\vertex (f) at ( 0.65, 1.65) {g};
		\diagram* {
			(a) -- [fermion] (m) -- [fermion] (c),
			(e) -- [gluon]   (f),
			(b) -- [double, solid] (m) -- [double, solid] (d),
		};
		\end{feynman}
		\end{tikzpicture}
		\caption{Representative Feynman diagram of mono-jet production along with DM.}
		\label{fig:monojet}
	\end{figure}
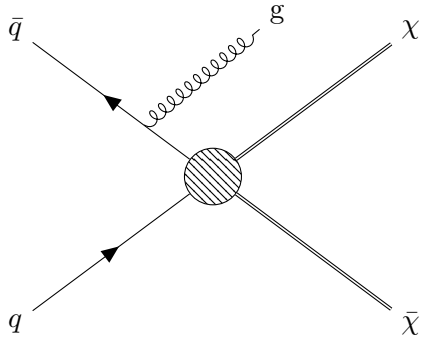

	Any non-vanishing effective interactions between DM and quarks, listed in Table~\ref{tab:operators}, should pair-produce DM through $p$-$p$ collision at the LHC, if its mass is within the reach of LHC energy, $\sqrt{s}=13$ TeV. However, production of solely DM particles does not yield any detectable signals within the detector, since they do not interact with the detector.  Only if the DM particles are accompanied by some visible particles, then the imbalance of the transverse momentum ($p_T$) can be measured by detecting the visible particles, such as mono-jet, where one extra jet is  produced along with the DM (see Fig.~\ref{fig:monojet} for a representative Feynman diagram), and results in signals of energetic jets with missing $p_T$ ($\MET$),
	\br
	 pp\to \chi\bar{\chi}j\to \MET j.
	\er
	
	At the LHC Run-II experiment, both ATLAS and CMS experiments have reported their results of mono-jet searches~\cite{ATLAS:2021kxv,CMS:2021far}. Ref.~\cite{ATLAS:2021kxv} looks for jets with high $p_T$ with different selection criteria on $p_T$ of the jet. Since the mono-jet cross-section drops with the $p_T$ of jet, a lower $p_T$ selection includes a larger fraction of events in the low $ p_T$ region, but fewer events in the high $ p_T$ region. Contrarily, a higher $ p_T$ cut includes the fraction of events from the higher $ p_T$ phase space only. As the scaling of the cross-section of SM backgrounds with $ p_T$ may not necessarily be identical to the DM signal, mono-jet analyses with different $ p_T$ cuts result in different levels of exclusion on the DM-quark interactions. In order to interpret results of mono-jet searches within the framework of effective interactions considered here, we closely follow analysis strategy described in Ref.~\cite{ATLAS:2021kxv}, as shown in Table~\ref{tab:monojet-data}.
    \begin{table}
		\caption{\small 95$\%$ CL upper limits on the visible mono-jet production cross-section at the LHC ($\sqrt{s}=13$ TeV)~\cite{ATLAS:2021kxv}.}
		\begin{center}
			\begin{tabular}{ c c  }
				\hline
				\hline \\ [-2.5ex]
				Selection criteria & $\sigma^{95}$(fb) \\[0.5ex]
                \hline 
                \hline\\ [-2.5ex]
                $ p_T^j>200~\rm GeV$ &  736 \\
                $ p_T^j>1200~\rm GeV$ & 0.3\\ [0.25ex]
				\hline
				\hline
			\end{tabular}
		\end{center}
		\label{tab:monojet-data}
	\end{table}

     The hard-scattering matrix elements are generated in \nolinkurl{MG5aMC_atNLO-3.3.0}~\cite{Alwall:2014hca} using \nolinkurl{Feynrules} UFO~\cite{Alloul:2013bka} package, which is required to include the effective interactions for a given set of parameters ($\Lambda,m_\chi$). Events are produced up to one jet in Madgraph, and MLM-matching is used to avoid double counting, which is essential in reproducing a mono-jet analysis. The showering and hadronization are performed using \nolinkurl{PYTHIA8}~\cite{Sjostrand:2006za, Sjostrand:2007gs}.  Unlike the previous study~\cite{March-Russell:2012elz}, we consider the ATLAS detector effects using \nolinkurl{Delphes}~\cite{deFavereau:2013fsa} ATLAS-card. After applying specific cuts on $p_T^j$ (see Table~\ref{tab:monojet-data}), we can determine the acceptance rate of the signal. It is obtained by dividing the number of events that meet the criteria on $p_T^j$ ($N$) by the total number of simulated events ($N_0$), i.e., $\mathcal{A}=N/N_0$. However, it is important to note that various detector efficiencies ($\epsilon$) are already applied during the simulation, and their impacts are already included in the value of $N$. Therefore, for a given parameter space, taking into account the effects of DM-EFT, we calculate the measurable cross-section, $\sigma_{\rm mes}=\sigma\times\mathcal{A}$. If $\sigma_{\rm mes}$ is greater than the experimentally determined $95\%$ upper limit, then we claim that the given choice of \{$\Lambda,m_\chi$\} is excluded. Repeating this procedure for various values of \{$\Lambda,m_\chi$\}, finally we obtain the excluded region in the $\Lambda-m_\chi$ plane due to the mono-jet searches, as shown in Figs.~\ref{fig:bounds_SI}-\ref{fig:bounds_5}.
     
 	\subsection{Direct detection experiments}
	DM can interact with the nucleon via the effective interactions listed in the Table~\ref{tab:operators}. The strength (1/$\Lambda$) of the interactions affect the DM-nucleon scattering cross-section. Thus, the bounds on the DM-nucleon scattering cross-section can be recast as constraints on these interactions in terms of $\Lambda$ and mass of DM ($m_\chi$). Notice that, the spin-dependence of the scattering cross-section significantly impacts the bounds placed on each effective operator in Table \ref{tab:operators}. Some of the interactions get suppressed by the powers of DM velocity ($v$). For example, the vector current singles out the temporal component of a spinor and the axial-vector current picks up the spatial component. Consequently, the combinations of these two ($\mathcal{O}_{av}$ or $\mathcal{O}_{va}$) are velocity suppressed, and eventually leading to weaker bounds from the DD experiments. On the other hand, spin-independent DM-nucleon scattering cross-section often turn out to be restricted severely. Considering universal coupling to quarks, the scattering cross-section for each of the non-suppressed spin-independent interactions listed in Table~\ref{tab:operators} are given by \cite{March-Russell:2012elz},
	\br
	\sigma_{\rm SI}^{\mathcal{O}_{ss}}\sim \frac{1}{\pi \Lambda^4}\mu_p^2f_p^2,\\
	\sigma_{\rm SI}^{\mathcal{O}_{vv}}\sim \frac{9}{\pi\Lambda^4}\mu_p^2,
	\label{eq:dd1}
	\er
	where, $\mu_p={m_{\chi}m_p}/{(m_{\chi}+m_p)}$ is the reduced mass of the nucleon-DM system and $f_p$ is the DM effective coupling to protons. Here we use $f_p=0.3$~\cite{Belanger:2008sj}. The non-suppressed scattering cross-section for the spin-dependent operators listed in Table~\ref{tab:operators} is \cite{March-Russell:2012elz}:
	\br 
	\sigma_{\rm SD}^{\mathcal{O}_{tt}}\sim 4\times\sigma_{\rm SD}^{\mathcal{O}_{aa}}\sim \frac{16}{\pi\Lambda^4}\mu_p^2\left(\sum_{q}\Delta_q^p\right)^2\,,
	\label{eq:dd2}
	\er
	where $\Delta_q^p$ accounts for the spin content of the nucleon and $(\sum_{q}\Delta_q^p)^2\simeq 0.32$ \cite{COMPASS:2007esq}.
	
The LUX-ZEPLIN (LZ) experiment currently provides the most stringent constraint on the cross-section for SI scattering between DM and nucleons. Specifically, they reported a limit of $\rm 6.5 \times 10^{-48}~ cm^2$ on $\sigma_{\rm SI}$ at 90$\%$ CL for 19 GeV DM mass~\cite{LZ:2022ufs}. In parallel, the DarkSide-50 experiment excluded a cross-section of $\rm 10^{-41}~cm^2$ for DM-nucleon SI interactions with a 90$\%$ CL, for a DM particle mass of 1.8 GeV \cite{Agnes:2018ves}.
The constraints on the SD scattering cross-sections of dark matter with protons and neutrons have also improved for large ranges of DM particle masses. Notably, the most stringent limits on DM-neutron interactions originate from the XENON-$n$T experiment~\cite{Aprile:2019dbj}, having a minimum of $6.3\times 10^{-42}\rm cm^2$ for DM mass 30 GeV at 90$\%$ CL. Whereas, for DM-proton interactions the PICO-60 experiment \cite{Amole:2019fdf} sets the strongest constraints, e.g., $2.5\times 10^{-41}\rm cm^2$ for DM mass 25 GeV at 90$\%$ CL.

To interpret these findings in the EFT context, we have converted the limits on DM-nucleon scattering cross-sections to the plane defined by the dark matter particle mass ($m_\chi$) and the EFT scale ($\Lambda$), using Eqns..\,(\ref{eq:dd1}) and (\ref{eq:dd2}). Additionally, for the sake of comparison against the constraints outlined in Ref.~\cite{March-Russell:2012elz}, we have included the bounds predicted by the XENON-100 experiment~\cite{XENON100:2011uwh}. These comparisons are illustrated in Figs.\,\ref{fig:bounds_SI} and \ref{fig:bounds_SD}.    

    \subsection{Constraints on the $\Lambda-m_{\chi}$ plane}

\begin{figure}
	\begin{subfigure}[b]{0.5\textwidth}
		\centering
		\includegraphics[width=9.2 cm]{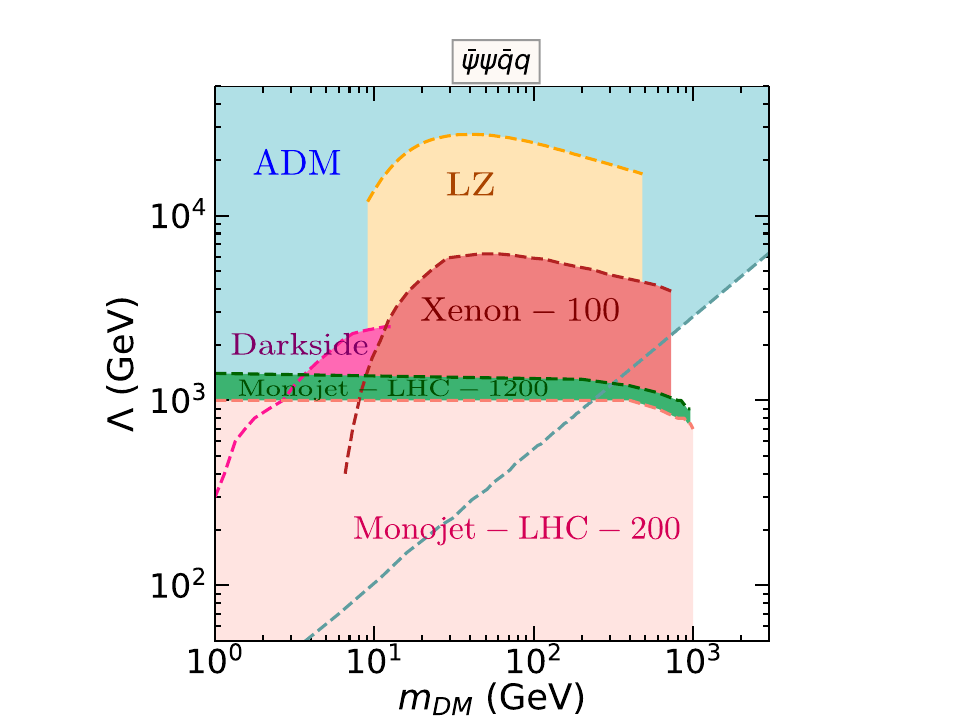}
	\end{subfigure}
	\begin{subfigure}[b]{0.5\textwidth}
		\centering
		\includegraphics[width=9.2 cm]{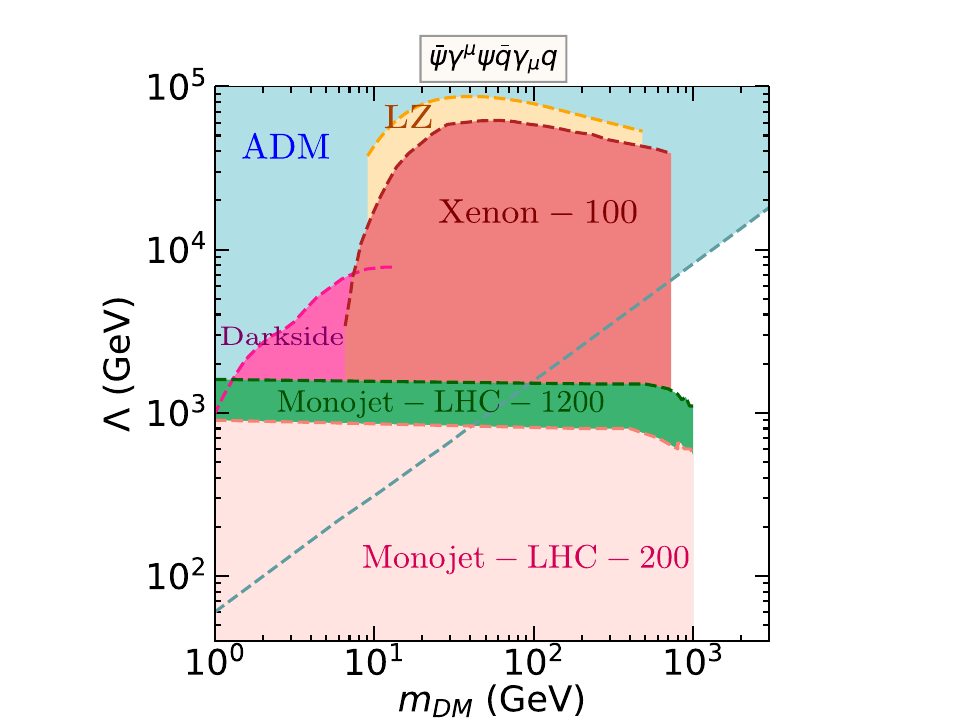}
	\end{subfigure}
	\caption{Current exclusion limits on the scale $\Lambda$ at different DM mass ($ m_{DM}$) for the scalar ($\mathcal{O}_{ss}\equiv \frac{1}{\Lambda^2}\bar{\psi}\psi\bar{q}q$)\,(left) and vector ($\mathcal{O}_{vv}\equiv\frac{1}{\Lambda^2}\bar{\psi}\gamma^{\mu}\psi\bar{q}\gamma_{\mu}q$) (right) type interaction of DM with quarks from different experimental observations and ADM considerations, labeled on the respective regions with darker shades. The experimental exclusions extends up to the bottom of the plots and overlapping regions are implicit.}
	\label{fig:bounds_SI}
\end{figure}

\begin{figure}
	\begin{subfigure}[b]{0.5\textwidth}
		\centering
		\includegraphics[width=9.2 cm]{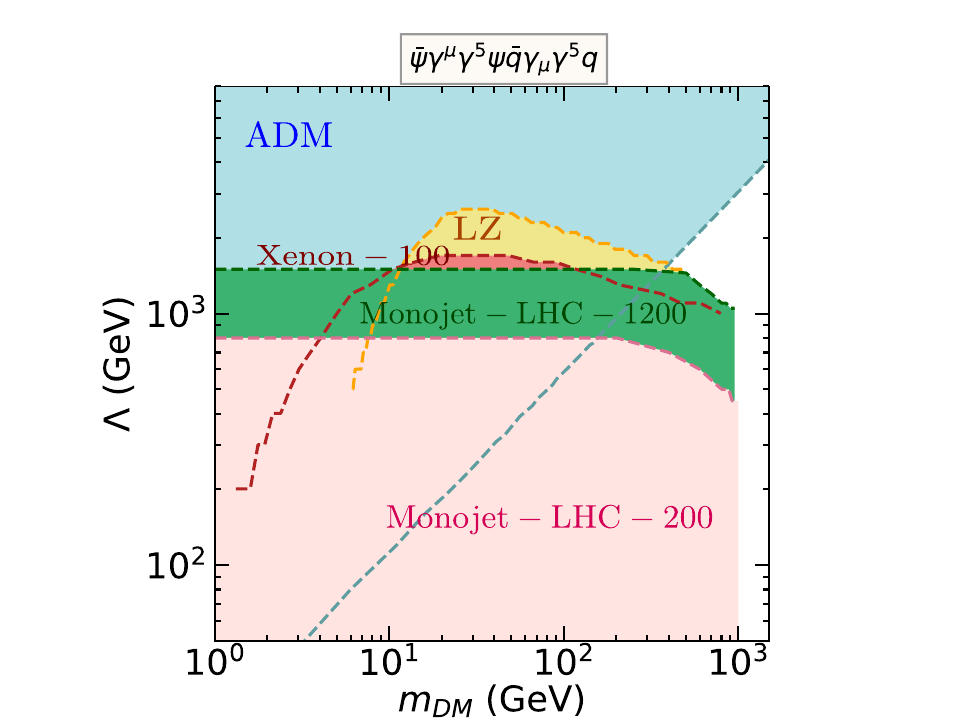}
	\end{subfigure}
	\begin{subfigure}[b]{0.5\textwidth}
		\centering
		\includegraphics[width=9.2 cm]{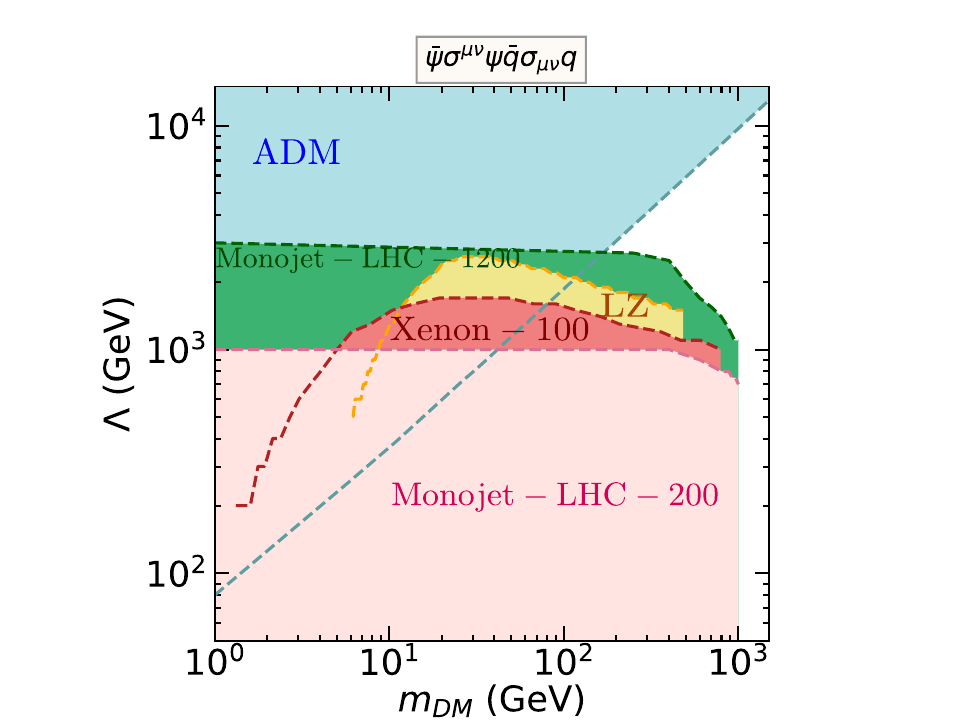}
	\end{subfigure}
	\caption{Same as Fig.~\ref{fig:bounds_SI} but for axial-vector ($\mathcal{O}_{aa}\equiv\frac{1}{\Lambda^2}\bar{\psi}\gamma^{\mu}\gamma^5\psi\bar{q}\gamma_{\mu}\gamma^5q$ ) (left) and tensor ($\mathcal{O}_{tt}\equiv\frac{1}{\Lambda^2}\bar{\psi}\sigma^{\mu\nu}\psi\bar{q}\sigma_{\mu\nu}q$) (right) type interaction of DM with quarks.}
	\label{fig:bounds_SD}
\end{figure}

\begin{figure}
	\begin{subfigure}[b]{0.5\textwidth}
		\centering
		\includegraphics[width=9.2 cm]{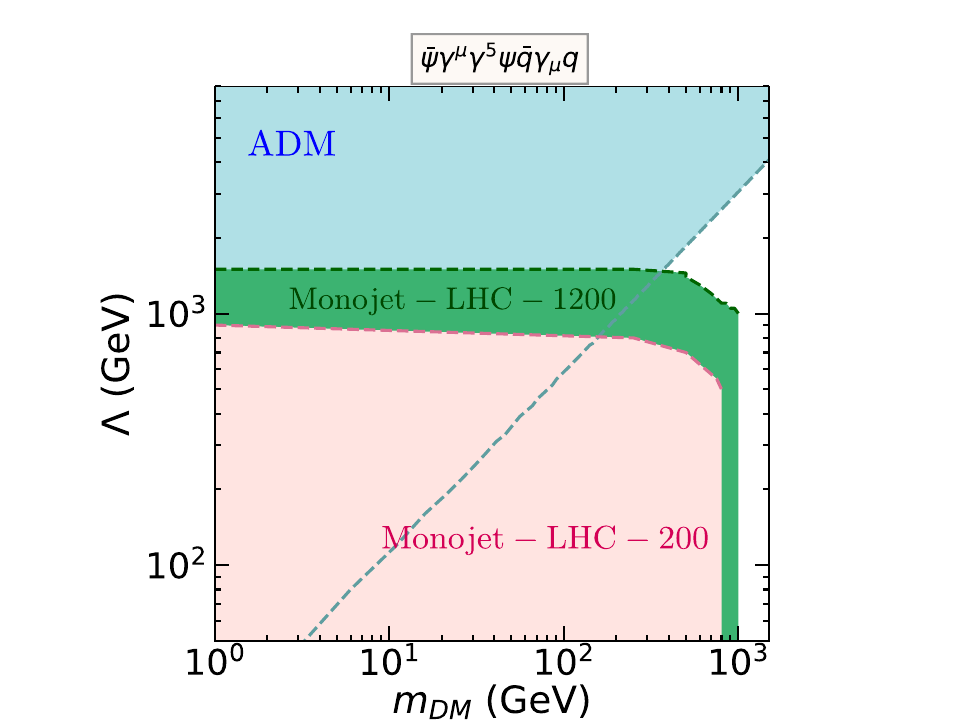}
	\end{subfigure}
	\begin{subfigure}[b]{0.5\textwidth}
		\centering
		\includegraphics[width=9.2 cm]{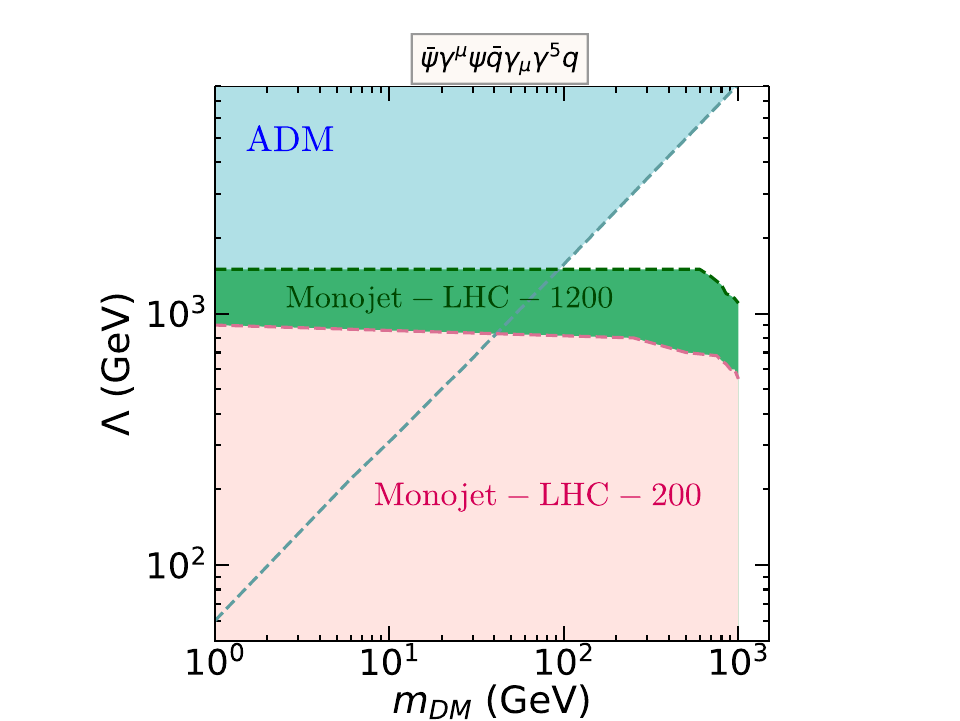}
	\end{subfigure}
	\caption{Same as Fig.~\ref{fig:bounds_SI} but for {axial-vector$-$vector ($\mathcal{O}_{av}\equiv \frac{1}{\Lambda^2}\bar{\psi}\gamma^{\mu}\gamma^5\psi\bar{q}\gamma_{\mu}q$)(left) and vector$-$axial-vector ($\mathcal{O}_{va}\equiv$ $\frac{1}{\Lambda^2}\bar{\psi}\gamma^{\mu}\psi\bar{q}\gamma_{\mu}\gamma^5q$) (right)} type interaction of DM with quarks.}
	\label{fig:bounds_3}
\end{figure}

\begin{figure}
	\begin{subfigure}[b]{0.5\textwidth}
		\centering
		\includegraphics[width=9.2 cm]{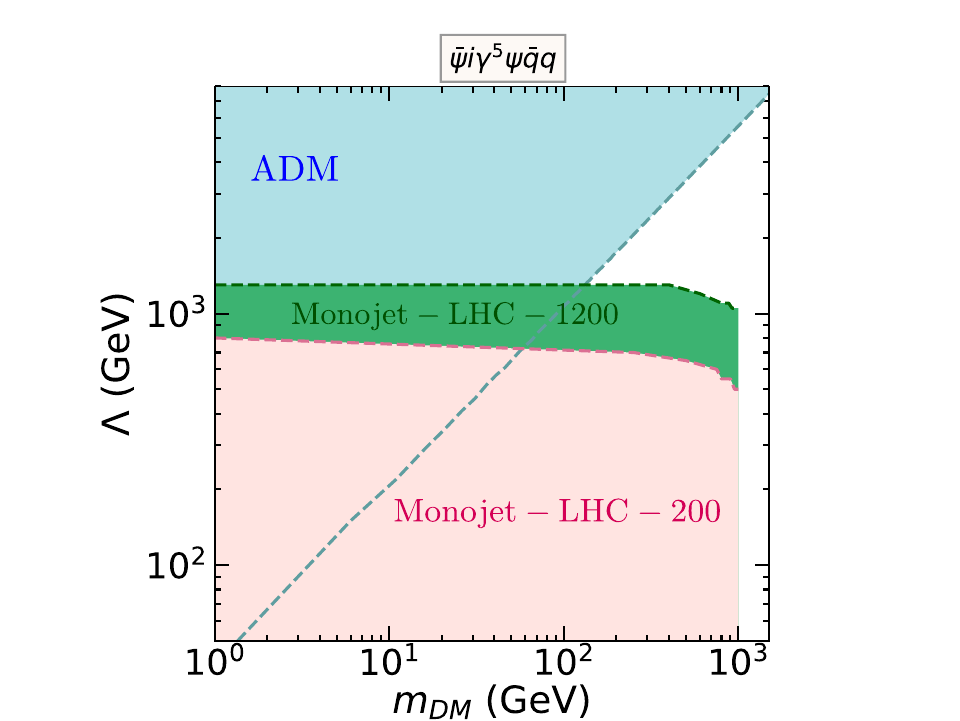}
	\end{subfigure}
	\begin{subfigure}[b]{0.5\textwidth}
		\centering
	\includegraphics[width=9.2 cm]{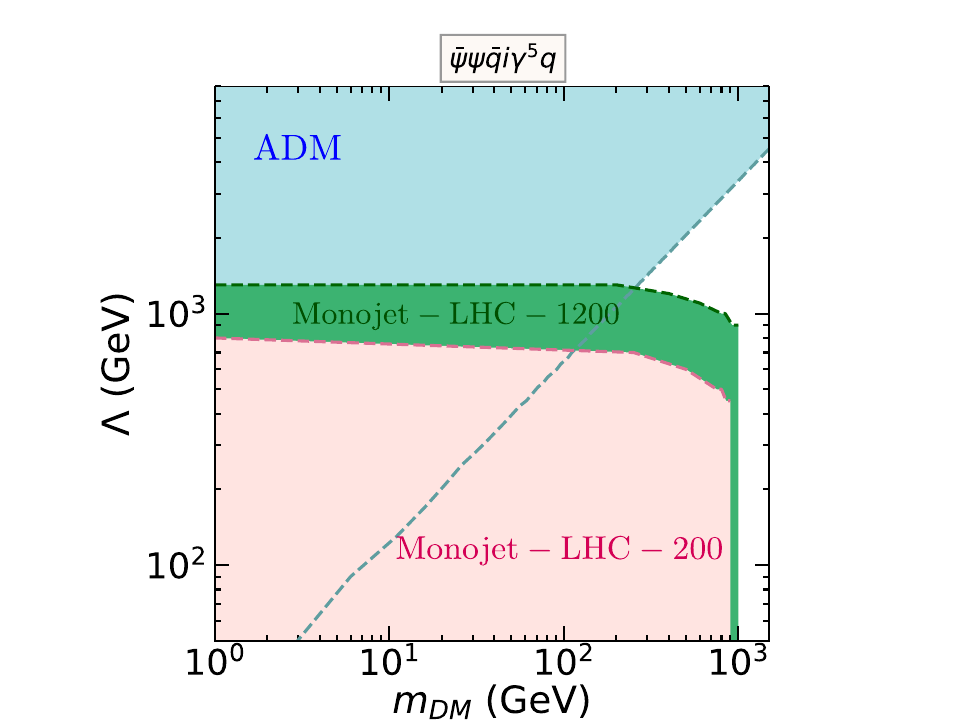}
	\end{subfigure}
	\caption{Same as Fig.~\ref{fig:bounds_SI} but for {pseudo-scalar$-$scalar ($\mathcal{O}_{ps}\equiv$ $ \frac{1}{\Lambda^2}\bar{\psi}i\gamma^5\psi\bar{q}q$ ) (left) and scalar$-$pseudo-scalar ($\mathcal{O}_{sp}\equiv$ $ \frac{1}{\Lambda^2}\bar{\psi}\psi\bar{q}i\gamma^5q$ ) (right)} type interaction of DM with quarks.}
	\label{fig:bounds_4}
\end{figure}

\begin{figure}
	\begin{subfigure}[b]{0.5\textwidth}
		\centering
		\includegraphics[width=9.2 cm]{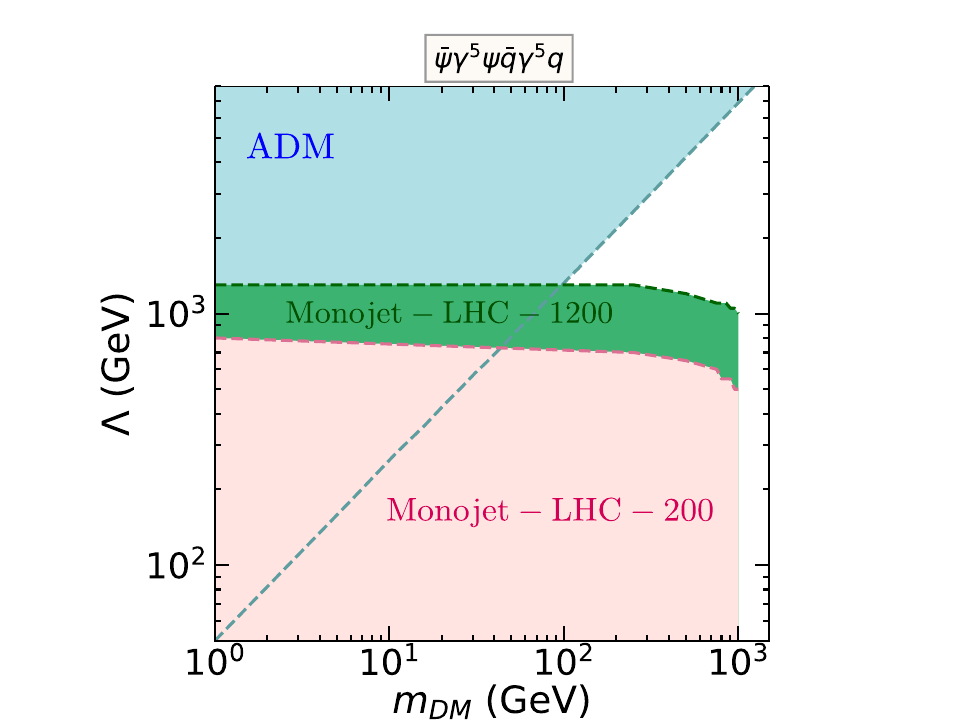}
	\end{subfigure}
	\begin{subfigure}[b]{0.5\textwidth}
		\centering
		\includegraphics[width=9.2 cm]{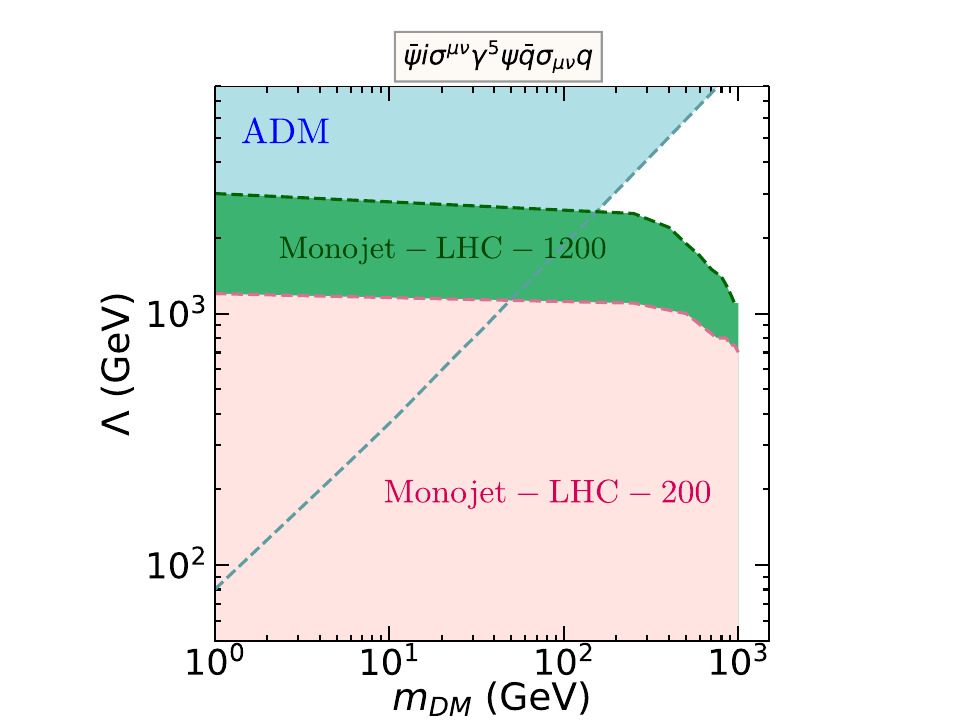}
	\end{subfigure}
	\caption{Same as Fig.~\ref{fig:bounds_SI} but for {pseudo-scalar$-$pseudo-scalar ($\mathcal{O}_{pp}\equiv$$\frac{1}{\Lambda^2}\bar{\psi}\gamma^5\psi\bar{q}\gamma^5q$) (left) and pseudo-tensor$-$tensor ($\mathcal{O}_{pt}\equiv$$\frac{1}{\Lambda^2}\bar{\psi}i\sigma^{\mu\nu}\gamma^5\psi\bar{q}\sigma_{\mu\nu}q$) (right)} type interaction of DM with quarks.}
	\label{fig:bounds_5}
\end{figure}

In Figs.~\ref{fig:bounds_SI}-\ref{fig:bounds_5}, the exclusion on $\Lambda$ as a function of ADM mass are shown along with the results due to various DD experiments, such as LZ, Xenon, Darkside. The bounds from the mono-jet searches are presented for two jet-$p_T$ selection criteria, $p_T(j)>200$ GeV (LHC-200) and $p_T(j)>1200$ GeV (LHC-1200). The current DD exclusions are also presented in the same plot to gather an idea of the overall picture of the exclusions.

Clearly, the current DD limits are found to be much stronger compared to those presented in the Ref.~\cite{March-Russell:2012elz}. For large part of the mass region, the DD bounds dominate the exclusion on the DM-quark interactions. However, in low DM-mass regions, even in case of the non-suppressed interactions, we found that the bounds coming from the mono-jet searches supersede the DD limits. Especially for the $\mathcal{O}_{aa}$ and $\mathcal{O}_{tt}$ operator, the current limits exclude a good fraction of region. Indeed, the DD measurements are found to be not sensitive for operators with suppressed interactions. In these cases the mono-jet searches put the dominant constraints for these interactions (see Figs.~\ref{fig:bounds_3}-\ref{fig:bounds_5}).

\section{Constraining ADM-Lepton Interactions}
In this section we consider the case of leptophilic DM. Note in these models DM-quark interactions occur only at the loop level. Consequently, the limits obtained based on the mono-jet searches at hadron colliders and nuclear recoil dependent DD experiments become much weaker in this case. Moreover, the LHC being a hadron collider, the production of DM, which is not coupled to quarks or gluons, is suppressed. On the other hand, leptophilic DM attracts bounds from monophoton production at lepton colliders, such as LEP, or may be searched at future lepton colliders, e.g., ILC, FCC-ee. Interestingly, studies of effects of DM capture in compact stars help constrain DM-lepton interactions strongly, which we will discuss in Section~\ref{DD_stars}.

\subsection{Mono-photon searches at LEP}
\label{sec:LEP}
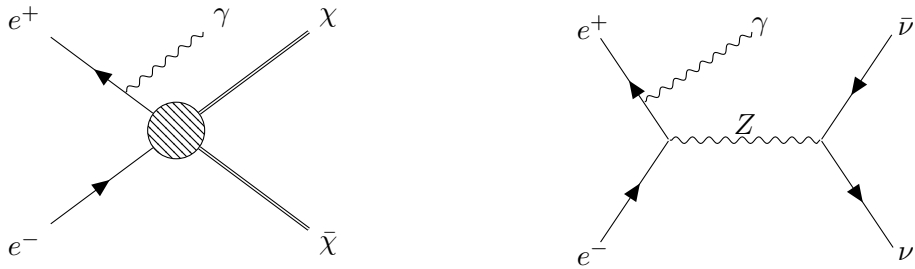
\begin{figure}
	\centering
	\begin{subfigure}[b]{0.49\textwidth}
		\centering
		\begin{tikzpicture}
		\begin{feynman}
		\vertex[blob] (m) at ( 0, 0) {\contour{white}{}};
		\vertex (a) at (-2,-1.5) {$e^-$};
		\vertex (b) at ( 2,-1.5) {$\bar{\chi}$};
		\vertex (c) at (-2, 1.5) {$e^+$};
		\vertex (d) at ( 2, 1.5) {$\chi$};
		\vertex (e) at (-0.8, 0.4) {};
		\vertex (f) at ( 0.6, 1.5) {$\gamma$};
		\diagram* {
			(a) -- [fermion] (m) -- [fermion] (c),
			(e) -- [photon]   (f),
			(b) -- [double, solid] (m) -- [double, solid] (d),
		};
		\end{feynman}
		\end{tikzpicture}
	\end{subfigure}
	\begin{subfigure}[b]{0.49\textwidth}
		\centering
		\begin{tikzpicture}[every node/.style={inner sep=0,outer sep=0}, line cap=rect]
		\begin{feynman}
		\vertex (a) at (-2,-1.5) {$e^-$};
		\vertex (b) at ( 2,-1.5) {$~\;\nu$};
		\vertex (c) at (-2, 1.5) {$e^+$};
		\vertex (d) at ( 2, 1.5) {$~\;\bar{\nu}$};

		\node (e) at (-1, 0  ) {};
		\node (f) at (+1, 0) {};
		\vertex (g) at (-1.33, 0.5) {};
		\vertex (h) at ( 0.2, 1.5) {$\gamma$};
		
		\diagram*{	
			(a) -- [fermion] (e) -- [fermion] (c),
			(d) -- [fermion] (f) -- [fermion] (b),
			(e) -- [photon, edge label=${Z_{}}_{}$] (f),
			(g) -- [photon, edge label={}] (h)
		};
		\end{feynman}
		\end{tikzpicture}
	\end{subfigure}
	\caption{(Left) Feynman diagram of DM production from $e^+e^-$ through effective interaction with additional photon from ISR; (right) Feynman diagram for the SM mono-photon production process.}
	\label{fig:feyn-diag}
\end{figure}

Similar to the mono-jet production in association with DM at the LHC, in an electron-position collider, e.g., at LEP, DM can be produced accompanied by a photon radiated off from the electron or positron (see Fig.~\ref{fig:feyn-diag} (left)), namely,
\br
 e^+e^-\to \chi\bar{\chi}\gamma\to \MET \gamma.
 \label{eq:monophoton_DM}
 \er
The SM counterpart of it involves production of neutrinos in place of DM (see Fig.~\ref{fig:feyn-diag} (right)), such as,
\br
   e^+e^-\to \nu\bar{\nu}\gamma\to \MET \gamma.
   \label{eq:monophoton_SM}
\er
 These processes give rise to a mono-photon signal at the detector, a potentially robust way to look for DM in $e^+e^-$ collider. In Ref.~\cite{Fox:2011fx}, the authors studied the mono-photon production in the $e^+e^-$ collider and predicted the excluded region in the $\Lambda-m_{\chi}$ plane for four effective operators (three of which are independent) of the DM-lepton interactions, considering the mono-photon analysis of the DELPHI experiment at LEP~\cite{DELPHI:2003dlq}, including a simulation of the effects of the DELPHI detector.

In this section, we revisit this LEP exclusions including all possible DM-lepton contact interactions (same as Table~\ref{tab:operators}, with $f=e$). Moreover, while presenting these LEP exclusions, we also include the limits from DD or indirect detection experiments, to review the overall picture of the current limits on leptophilic DM. Using these limits, we also infer the exclusions for the case of ADMs. This approach permits us to delineate the parameter spaces for viable ADM scenarios, similar to the analysis in the previous section.

In the following, we document our methodology to find the exclusions due to the LEP data. To start with, we performed a Monte-Carlo (MC) simulation of the SM mono-photon process, as in Eq.\,(\ref{eq:monophoton_SM}). This simulation is tailored to closely replicate the experimental data. We factor in all the relevant aspects such as detector efficiencies, resolutions, and effects of smearing.  Then we validated our MC simulation with the experimental data. Post-validation, we proceeded to employ the same configuration to generate events associated with dark matter for a wide range of parameter values, $\Lambda$ and $m_{\chi}$. Subsequently, we evaluated the degree of excess that could potentially manifest at LEP as a result of these dark matter events. Finally, this allows us to exclude regions in the $\Lambda$-$m_{\chi}$ plane.

We closely follow the original LEP analysis Ref.~\cite{DELPHI:2003dlq}, describing the measurements of the mono-photon events with the DELPHI detector corresponding to luminosity $\rm \mathcal{L}=650~pb^{-1}$. These measurements are presented in bins of,
\br
X_{\gamma}=\frac{E_{\gamma}}{E_{\rm beam}},
\er
where $E_\gamma$ and $E_{\rm beam}$ are the energies of the photon and the colliding beams respectively.
The DM events are generated using Madgraph5-aMC@NLO-3.3.0\cite{Alwall:2014hca} including the ISR photon, using a UFO model file for leptophilic DM produced using the Feynrules package~\cite{Alloul:2013bka}. Finally, mono-photon events are passed through PYTHIA8~\cite{Sjostrand:2006za,Sjostrand:2007gs} for showering. It is important to emphasize here that accurate measurements of photons arising from initial state radiation (ISR) is crucial to model a mono-photon process. In our analysis, we ensure that all final state photons pass the requirements for identification with the DELPHI detector. For completeness and clarity, we provide a brief overview of the detector efficiencies and resolutions in the following.

\begin{table}[t]
	\centering
	\caption{\small Efficiencies and energy resolutions of various components of the DELPHI detector. Trigger and acceptance efficiencies increase linearly, with specified values at the given energies. Energy unis are in GeV.}
	\label{tab:delphi_eff_res}
	\begin{adjustbox}{max width=\textwidth} 
	\begin{tabular}{ccccccc}
		\hline
		\hline
		\multirow{2}{*}{Detector} & \multirow{2}{*}{Angular coverage} & Overall & Trigger & Acceptance & Angular & Energy\\
		& & efficiency & efficiency & efficiency & requirement & resolution\\
		
		\hline
		\hline
		\multirow{3}{*}{HPC}& &  & $52\%$ at $E_\gamma =6$;  & $41\%$ at $E_\gamma =6$; &  & \\
		
		& $45^{\circ} < \theta < 135^{\circ}$ & $X_\gamma>0.6$  & $72\%$ at $E_\gamma =30$;  & $78\%$ at $E_\gamma =80$, & -  & $0.043 \oplus 0.32/\sqrt{E}$  \\
		
		&  &  & $84\%$ at $E_\gamma =100$  & and above &  &  \\
		
		\hline
		
		\multirow{3}{*}{FEMC}& \multirow{2}{*}{$12^{\circ} < \theta < 32^{\circ}$, or} &  & $93\%$ at $E_\gamma =10$;  & $57\%$ at $E_\gamma =10$; &  & \\
		
		& \multirow{2}{*}{($148^{\circ} < \theta < 168^{\circ}$)}  & $X_\gamma>0.1$  & $100\%$ at $E_\gamma =15$;  & $75\%$ at $E_\gamma =100$; & $\theta>28^{\circ}-80^{\circ} X_\gamma$  & $0.03 \oplus 0.12/\sqrt{E} \oplus 0.11/E$  \\
		
		& &  & and above  & $\times$ const. factor $89\%$ &  &  \\
		\hline
			\multirow{2}{*}{STIC}& $3.8^{\circ} < \theta < 8^{\circ}$, or & \multirow{2}{*}{$X_\gamma>0.3$} & Const. $48\%$ & \multirow{2}{*}{-} & \multirow{2}{*}{$\theta >9.2^{\circ}-9^{\circ}X_{\gamma}$} & \multirow{2}{*}{$0.0152 \oplus 0.135/\sqrt{E}$} \\
		
		& ($172^{\circ} < \theta < 176.2^{\circ}$)&   & all over & &   &  \\
		\hline
		\hline
	\end{tabular}
\end{adjustbox}
\end{table}

The DELPHI detector had three parts, namely, High-Density Projection Chamber (HPC), Forward Electromagnetic Calorimeter (FEMC), and Small Angle Tile Calorimeter (STIC). We present the energy resolutions and different efficiencies of various sub-detector in Table~\ref{tab:delphi_eff_res}. There is an overall efficiency for each sub-detector components which depends on $X_\gamma$, followed by energy-dependent trigger and acceptance efficiencies. The trigger and acceptance efficiencies increase linearly with energy. The slope of the increase can be obtained using a few fixed values at certain energies, which are presented in Table~\ref{tab:delphi_eff_res}. To reduce the large background of radiative Bhabha events in FEMC, and due to beam-gas interactions in STIC, an energy-dependent angular requirement is imposed. The energy resolution adds an energy-dependent smearing for each of the photon that have passed all the selection criteria. Following Ref.~\cite{Fox:2011fx}, we have added a Lorentzian smearing with a width of 0.052$E_\gamma$ GeV, which is necessary to obtain a more realistic distribution.

\begin{figure}
	\centering
	\includegraphics[width=8.5 cm]{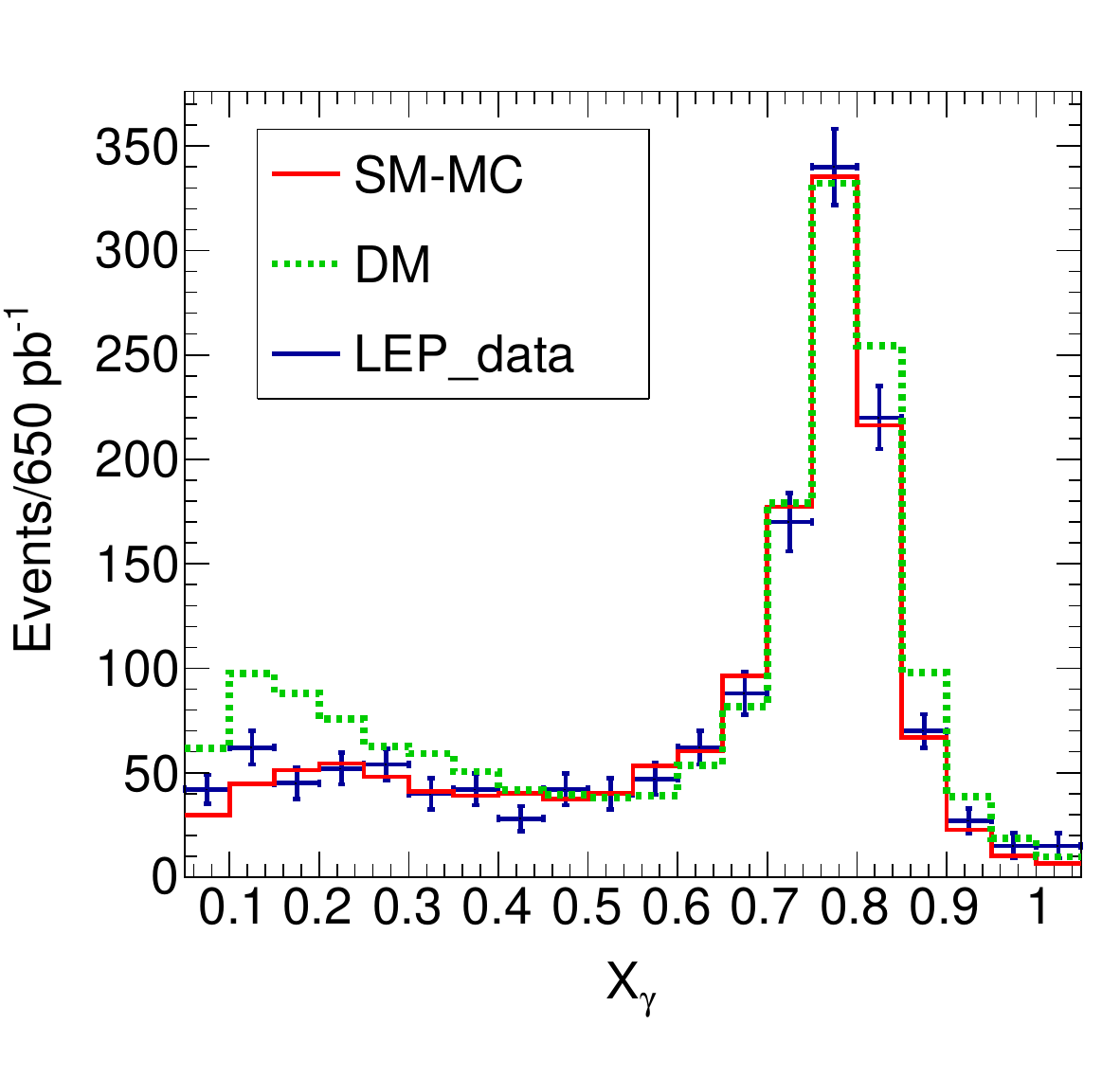}
	\caption{Distribution of $X_{\gamma}=E_{\gamma}/E_{\rm beam}$ for single photon event, where the agreement of the SM Monte-Carlo (red histogram) with the DELPHI-data (blue points with error bar) can be observed. The DM signal (green histogram), with vector-like interaction with $\rm \Lambda=300~GeV$ and $m_{\chi}=10~{\rm GeV}$, shows excess for the lower values of $X_{\gamma}$ and otherwise consistent with the SM.}
	\label{fig:data-mc}
\end{figure}

In order to demonstrate whether our simulated data correctly mimics the LEP measurements, we compare the p-value of $X_\gamma$ distribution corresponding to the SM mono-photon production (Eq.~\ref{eq:monophoton_SM}) with the DELPHI measurements. We obtain the data-MC agreement for $X_\gamma$ as shown in Fig.~\ref{fig:data-mc}. Some erroneous data can be found in the last bin, most likely due to inaccurate detector resolution modeling. We remove this bin from the $\rm \chi^2$-analysis. The $\rm \chi^2/dof$ is $\sim 14/19$ for 19 d.o.f corresponding to yields in 19 bins. A good agreement is found, confirming our simulation methodology.
In the same figure, the distribution for the DM signal is also plotted considering $ m_{\chi}=10$~GeV and $\rm \Lambda=300~GeV$, where the lower $X_{\gamma}$ bins show deviations due to DM effects. 

Now we use this setup to generate the mono-photon energy distribution for the mono-photon production for DM (Eq.~\ref{eq:monophoton_DM}) for each choice of $m_{\chi}$ and $\Lambda$. Calculation of $\chi^2$ between the photon energy distributions for SM-MC and the DM, it is possible to predict the allowed and ruled out DM parameter space. Results are shown for different operators in Figs.\,\ref{fig:exc1}-\ref{fig:exc5}. Notice that, these bounds are the strongest one for leptophilic dark matter in the $\lesssim 100$ GeV mass range, which are discussed later.

\subsection{Discovery potential of ADM at FCC-ee}
In this section we discuss the discovery potential of leptophilic DM in the FCC-ee (Future Circular Collider - electron-positron)  corresponding to the allowed region of parameter space. The FCC-ee is a proposed $e^+e^-$ collider with relatively higher center of mass (COM) energies and luminosities, measuring observables at unprecedented precision~\cite{FCC:2018evy}. The FCC-ee collider is proposed to operate in several stages, with different COM energies, such as $\sqrt{s}=91,160,240$ and 365 GeV. In our study, we have used the highest energy option ($\sqrt{s}=365$ GeV) to calculate the discovery reach of ADM scenarios. It is to be noted that, this sensitivity study is applicable to any leptophilic DM with contact interactions of Table~\ref{tab:operators}, and not limited to ADM.

The FCC-ee experiment should produce mono-photon events similar to LEP. However, the new detector set-up with increased COM energy is expected to probe wider ranges of parameters ($\Lambda, m_{\chi}$).

\begin{figure}
	\begin{subfigure}[b]{0.55\textwidth}
		\raggedleft
		\includegraphics[width=8.4 cm]{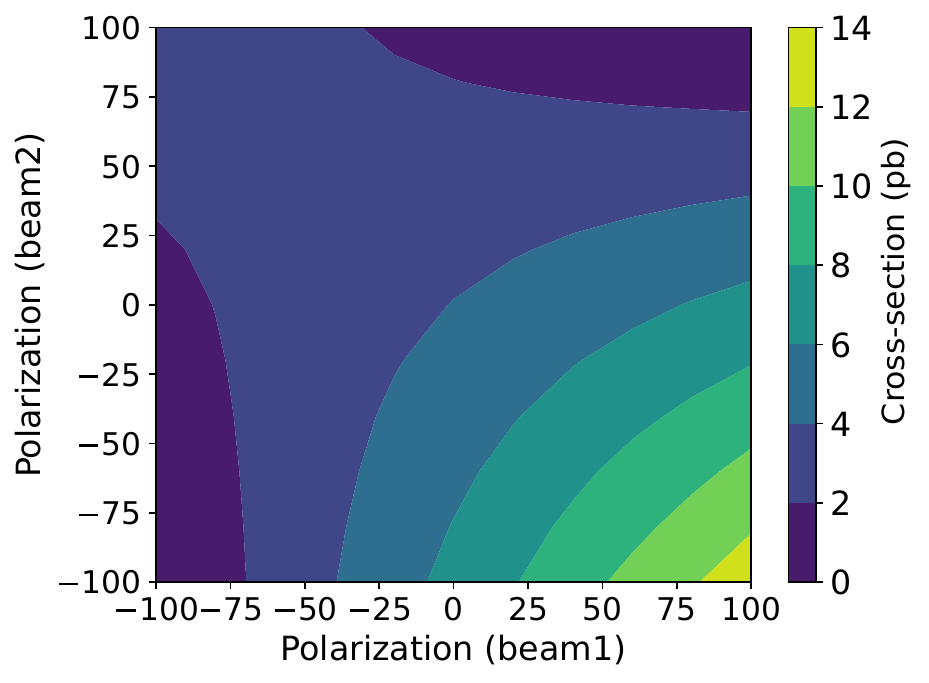}
	\end{subfigure}
	\begin{subfigure}[b]{0.5\textwidth}
		\raggedright
		\includegraphics[width=6.8 cm]{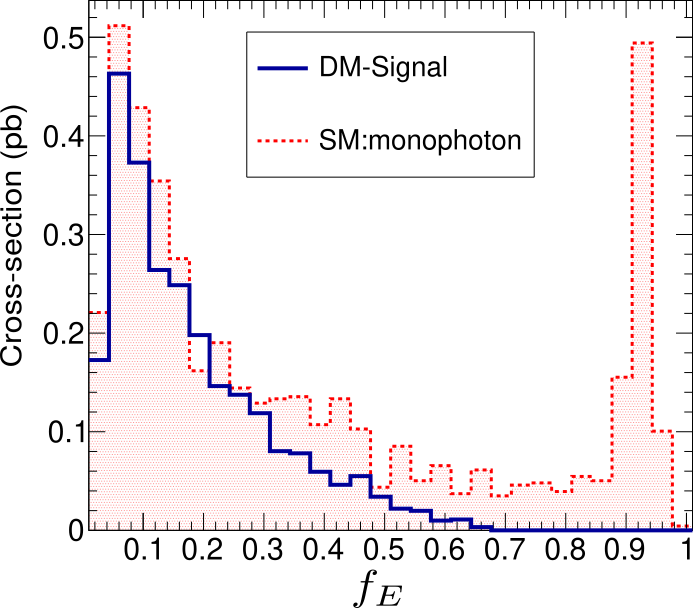}
	\end{subfigure}
	\caption{\small (Left) Variation of the SM mono-photon production cross-section ($e^+ e^- \to \nu\nu \gamma$) at FCC-ee with variation of beam-polarization; (right) distribution of the energy fraction of photon in the mono-photon events, for DM-signal (choosing $\rm \Lambda=400~GeV, m_\chi=100~GeV$, and vector-like interaction) and SM-background.}
	\label{fig:pol}
\end{figure}

 The production cross-sections of both the signal and backgrounds depend highly on the polarization of the $e^+$ and $e^-$ beams. For the sake of demonstration, we compute the cross-section of the SM mono-photon background using Madgraph5-aMC@NLO-3.4.1\cite{Alwall:2014hca}, varying the polarization of the beams. The variation of the cross-section with different choices of beam polarizations are presented in Fig.~\ref{fig:pol}\,(left). As the beam polarization setting of FCC-ee is not fixed yet, we used an unpolarized ($50\%$-L and $50\%$-R combination) beam for representative purposes.

For the signal, we essentially have one detectable object, namely the photon in the final state. All other kinematic variables such as $\rm \MET$ and the transverse mass, $M_{T}(\gamma,\MET)$, are strongly correlated to the energy of the photon. Hence a selection for photon energy results in the effective discrimination between signal and background. Hence, a selection as
\br
f_{E}=E_{\gamma}/E_{\rm beam}<0.3,
\er 
is found to be very powerful to suppress backgrounds as shown in Fig.~\ref{fig:pol}\,(right panel). The resulting signal significances, ${S}/{\sqrt{S+B}}$, are estimated at $\rm \mathcal{L}=340~fb^{-1}$ for each of the given parameter spaces $\{m_\chi,\Lambda\}$. Repeating this analysis for different choices of $\{m_\chi,\Lambda\}$, we present the region for which ${S}/{\sqrt{S+B}}>3.0$, i.e., the 3-$\sigma$ reach. The results are presented in the following subsection.

\subsection{Constraints from compact stars and direct detection}
\label{DD_stars}
As seen in the previous sections, lepton-colliders are sensitive to a relatively smaller range of leptophilic ADM masses. Comparatively, DM capture in compact objects can be used to set bounds on a wider range of DM mass~\cite{1985ApJ296679P,1987ApJ321560G,1987ApJ321571G,Bertone:2007ae,Bramante:2017xlb,Dasgupta:2019juq,Dasgupta:2020dik}. Especially, neutron stars (NS)~\cite{Kouvaris:2007ay,Kouvaris:2010vv,Baryakhtar:2017dbj,Bell:2018pkk,Chen:2018ohx,Acevedo:2019agu,Raj:2017wrv,Bramante:2017xlb,deLavallaz:2010wp,Bell:2019pyc, Bell:2021fye} and white dwarfs (WD)~\cite{Bertone:2007ae,McCullough:2010ai,Hooper:2010es,Acevedo:2019gre,Graham:2018efk,Bell:2021fye} constrains DM-lepton interactions strongly for a MeV to TeV range of DM mass. Due to their high density, NS can capture dark matter very efficiently and can lead to the heating of NS~\cite{Baryakhtar:2017dbj,Raj:2017wrv}. Since old, isolated NS can naturally cool to temperatures below 1000 K, this heating can be utilized to establish constraints on the strength of DM interactions. For instance, Refs.~\cite{Bell:2019pyc, Bell:2021fye} have calculated these limits in detail for DM-electron interactions. 

A comparable scenario can arise with white dwarfs (WDs) as well. The core of a WD is primarily composed of a degenerate electron gas. Consequently, if there is interaction between DM and electrons, it would likely be more pronounced at the core of the WD. The DM can then scatter of the electrons and can heat up the WD. Also, a DM-dense WD core may result in DM annihilation, another source of rising the temperature of WD. The presence of DM would then prevent the natural cooling of the old, isolated WDs. Conversely, the absence of anomalous cooling in WDs can lead to stringent constraints on the strength of the interaction between DM and electrons. Following this, Ref.~\cite{Bell:2021fye} calculated the limits on DM-lepton interactions from WDs. 

On the other hand, among different DD experiments, Damic~\cite{DAMIC-M:2022aks,DAMIC-M:2023gxo} and SuperCDMS~\cite{SuperCDMS:2018mne} put the strongest constraint on DM-lepton interactions for MeV to GeV range of DM mass. However, these DD bounds are generally not competitive in the preferred mass range for ADM. Nevertheless, the upcoming DAMIC-M-1kg-yr experiment is expected to have a promising sensitivity~\cite{DAMIC-M:2022aks}, and we consider this projected constraint for illustrative purposes. Finally, we present an overall picture of the constraints on DM-lepton interactions from all possible sources.

\subsection{Constraints and discovery potential for leptophilic ADM}
In this subsection, we discuss the constraints on the ADM-lepton interactions imposed by the experimental measurements, as discussed in section \ref{sec:LEP}. In order to achieve a comprehensive perspective on the global exclusions associated with these interactions, we also incorporate important constraints originating from both indirect and direct searches. These additional constraints, gathered from Refs.~\cite{Bell:2019pyc,Bell:2021fye}, are presented alongside the new collider-based bounds. It is worth noting that the constraints arising from dark matter-nucleon scattering experiments are relatively weak. This is because these interactions can only occur at the loop-level. Conversely, experiments focusing on dark matter-electron scattering probe a range of much lower dark matter masses, a region that is not highly motivated by the ADM hypothesis.

\begin{figure}
	\begin{subfigure}[b]{0.5\textwidth}
		\centering
		\includegraphics[width=9.1 cm]{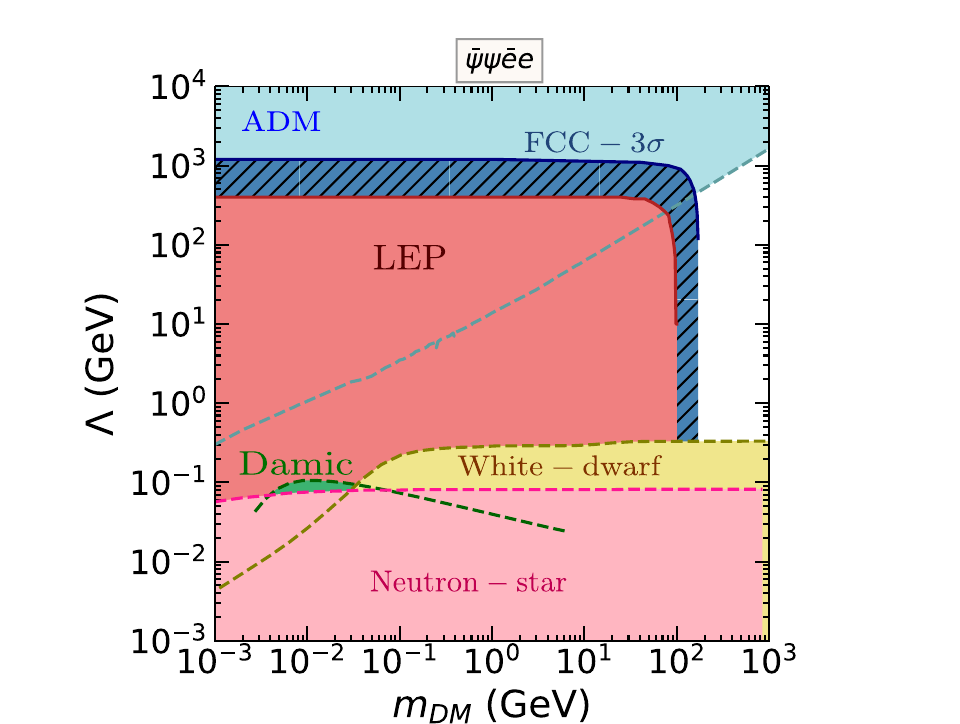}
	\end{subfigure}
	\begin{subfigure}[b]{0.5\textwidth}
		\centering
		\includegraphics[width=9.1 cm]{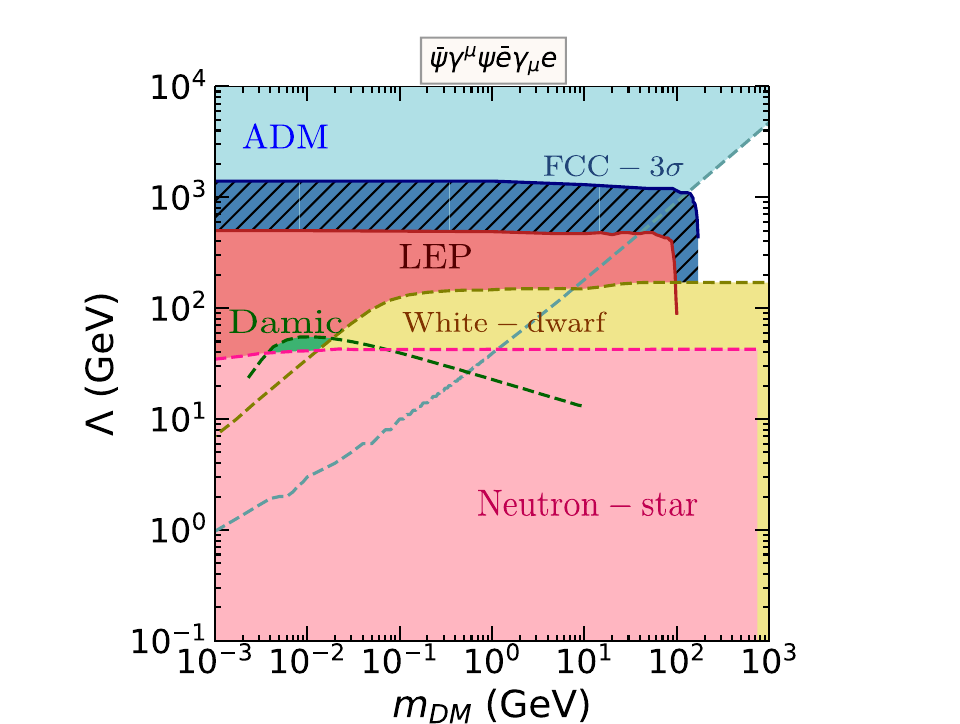}
	\end{subfigure}
	\caption{Limits on $\Lambda_e$ for {scalar ($\mathcal{O}_{ss}\equiv \frac{1}{\Lambda^2}\bar{\psi}\psi\bar{e}e$)(left) and vector ($\mathcal{O}_{vv}\equiv\frac{1}{\Lambda^2}\bar{\psi}\gamma^{\mu}\psi\bar{e}\gamma_{\mu}e$) (right)} type interaction of DM with electrons from different experimental observations and ADM scenario, labeled on the respective regions with darker shades. The striped dark blue region corresponds to the FCC-3$\sigma$ reach. The experimental exclusions/reach extends up to the bottom of the plots and overlapping regions are implicit.}
	\label{fig:exc1}
\end{figure}

\begin{figure}
	\begin{subfigure}[b]{0.5\textwidth}
		\centering
		\includegraphics[width=9.1 cm]{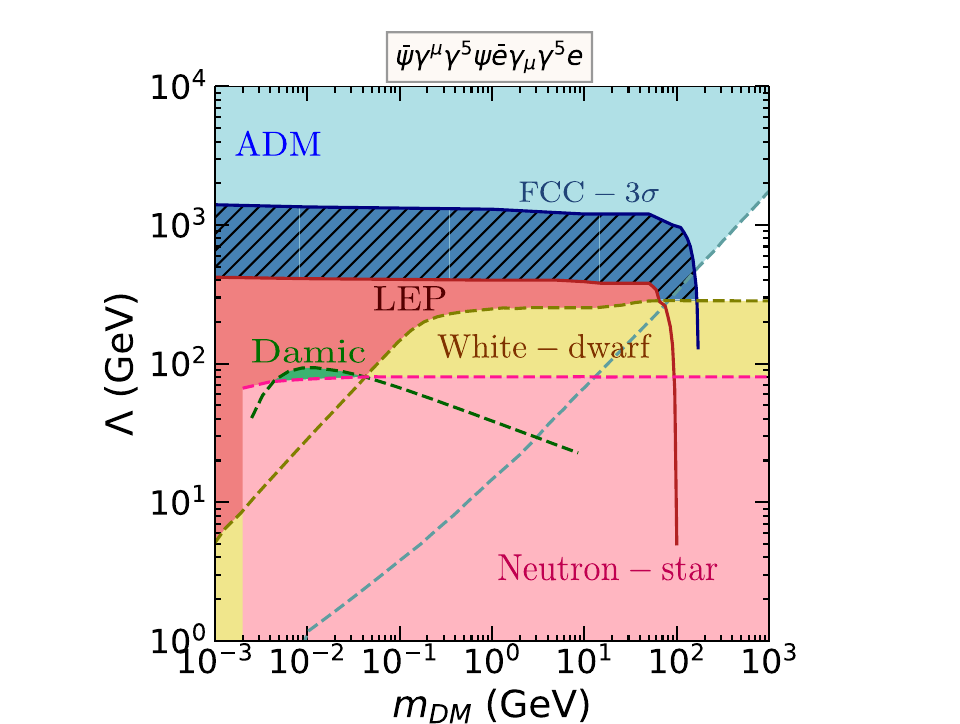}
	\end{subfigure}
	\begin{subfigure}[b]{0.5\textwidth}
		\centering
		\includegraphics[width=9.1 cm]{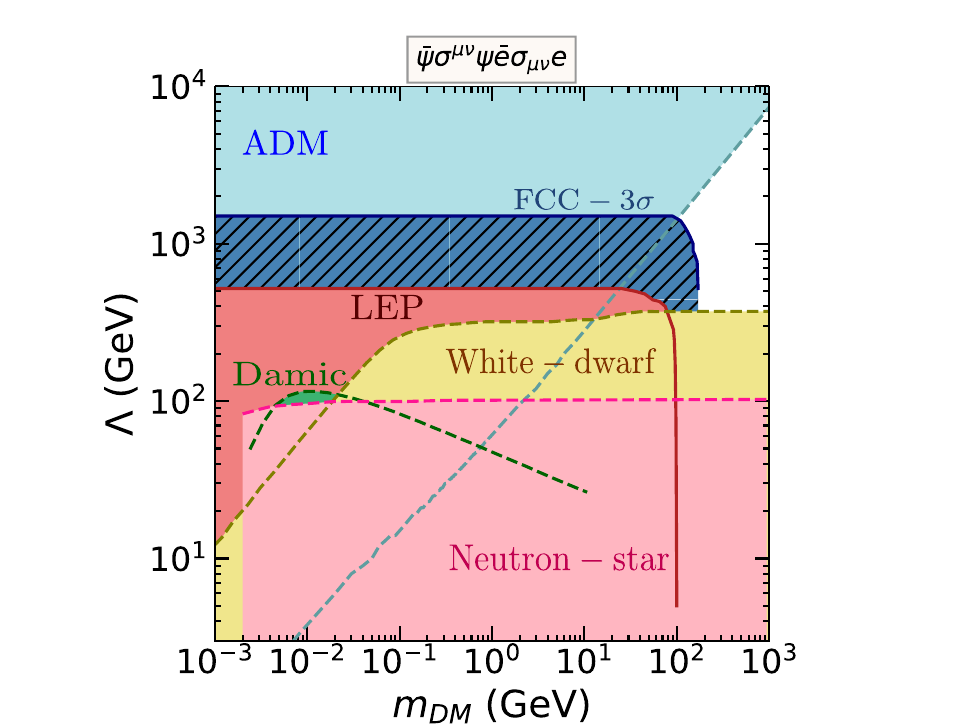}
	\end{subfigure}
	\caption{Same as Fig.~\ref{fig:exc1} but for {axial-vector ($\mathcal{O}_{aa}\equiv\frac{1}{\Lambda^2}\bar{\psi}\gamma^{\mu}\gamma^5\psi\bar{e}\gamma_{\mu}\gamma^5e$ ) (left) and tensor ($\mathcal{O}_{tt}\equiv\frac{1}{\Lambda^2}\bar{\psi}\sigma^{\mu\nu}\psi\bar{e}\sigma_{\mu\nu}e$) (right)} type interaction of DM with electrons.}
	\label{fig:exc2}
\end{figure}

\begin{figure}
	\begin{subfigure}[b]{0.5\textwidth}
		\centering
		\includegraphics[width=9.1 cm]{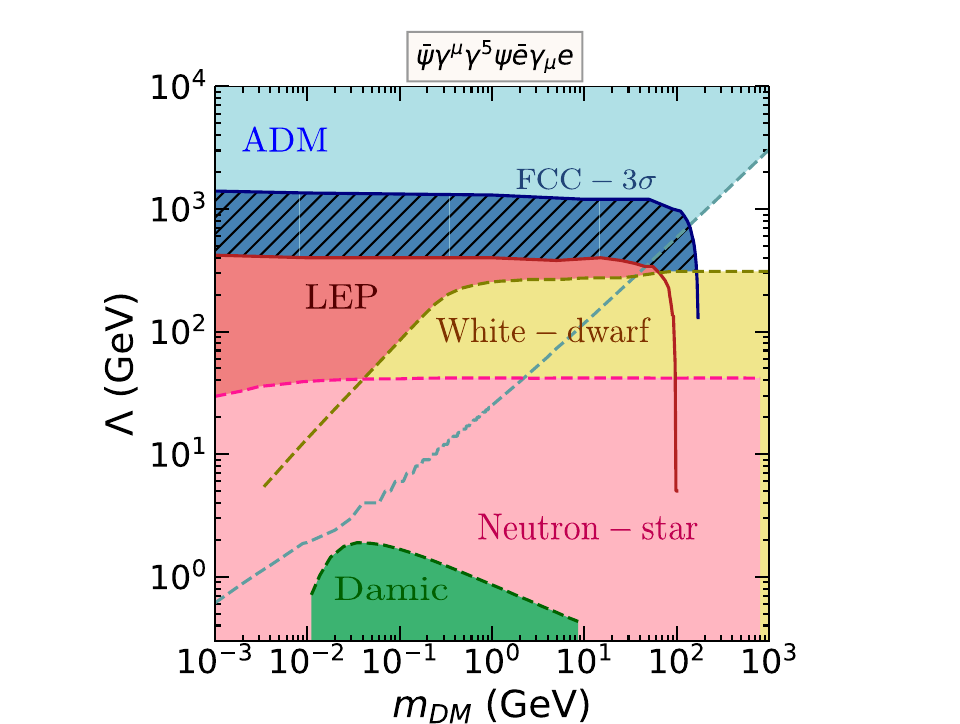}
	\end{subfigure}
	\begin{subfigure}[b]{0.5\textwidth}
		\centering
		\includegraphics[width=9.1 cm]{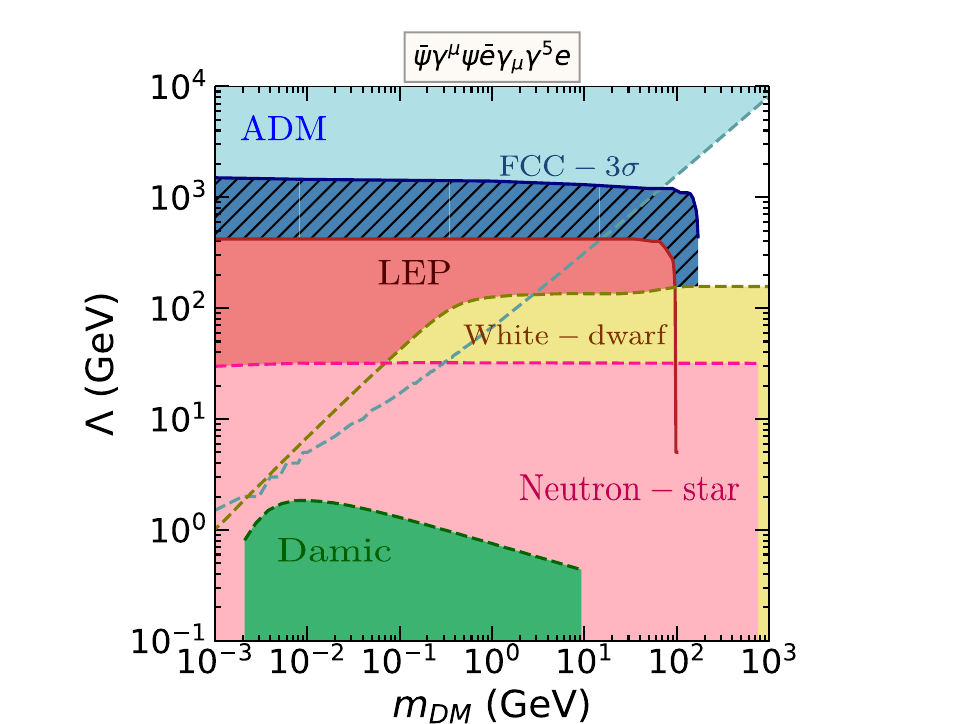}
	\end{subfigure}
	\caption{Same as Fig.~\ref{fig:exc1} but for {axial-vector$-$vector ($\mathcal{O}_{av}\equiv \frac{1}{\Lambda^2}\bar{\psi}\gamma^{\mu}\gamma^5\psi\bar{e}\gamma_{\mu}e$)\,(left) and vector$-$axial-vector ($\mathcal{O}_{va}\equiv$ $\frac{1}{\Lambda^2}\bar{\psi}\gamma^{\mu}\psi\bar{e}\gamma_{\mu}\gamma^5e$) (right)} type interaction of DM with electrons.}
	\label{fig:exc3}
\end{figure}

\begin{figure}
	\begin{subfigure}[b]{0.5\textwidth}
		\centering
		\includegraphics[width=9.1 cm]{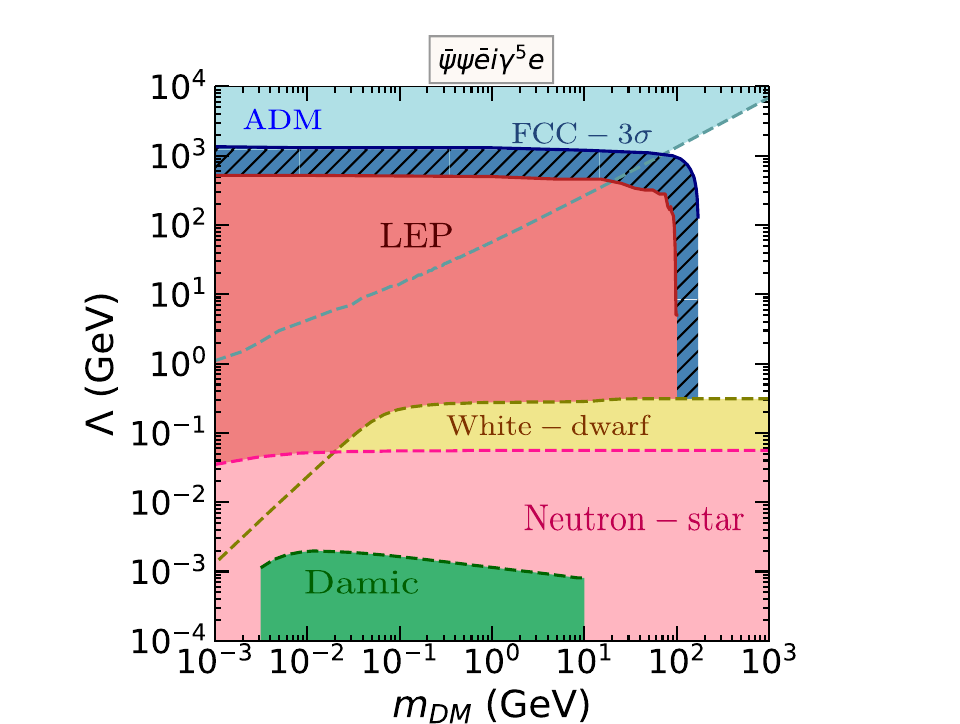}
	\end{subfigure}
	\begin{subfigure}[b]{0.5\textwidth}
		\centering
		\includegraphics[width=9.1 cm]{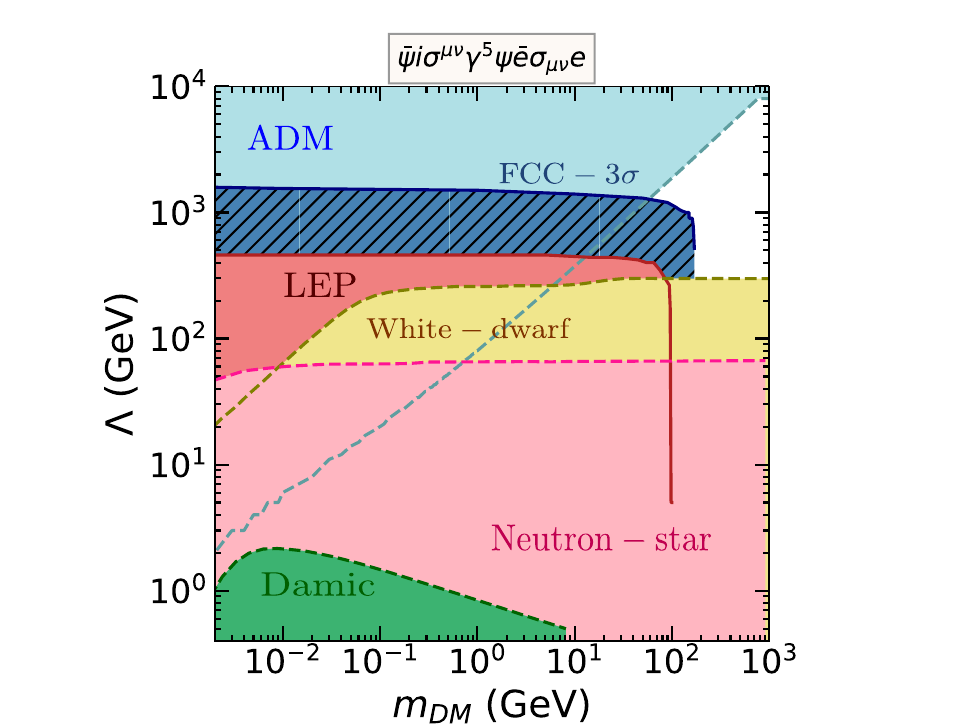}
	\end{subfigure}
	\caption{Same as Fig.~\ref{fig:exc1} but for {scalar$-$pseudo-scalar ($\mathcal{O}_{sp}\equiv$ $ \frac{1}{\Lambda^2}\bar{\psi}\psi\bar{e}i\gamma^5e$ ) (left) and pseudo-tensor$-$tensor ($\mathcal{O}_{pt}\equiv$$\frac{1}{\Lambda^2}\bar{\psi}i\sigma^{\mu\nu}\gamma^5\psi\bar{e}\sigma_{\mu\nu}e$) (right)} type interaction of DM with electrons.}
	\label{fig:exc4}
\end{figure}

\begin{figure}\vspace{-0.3cm}
	\begin{subfigure}[b]{0.5\textwidth}
		\centering
		\includegraphics[width=9.1 cm]{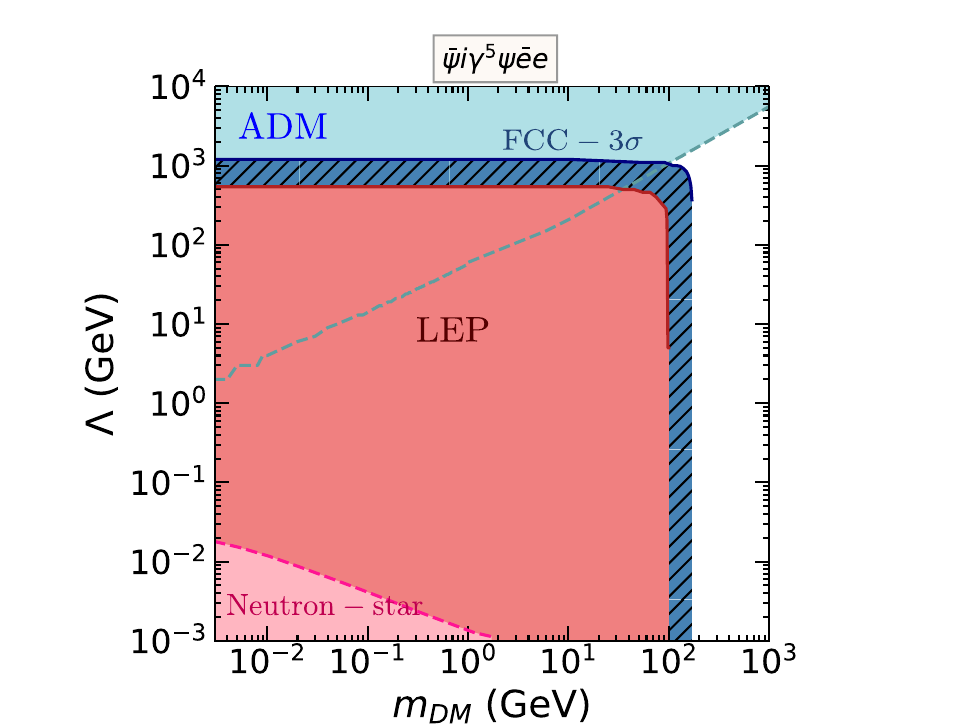}
	\end{subfigure}
	\begin{subfigure}[b]{0.5\textwidth}
		\centering
		\includegraphics[width=9.1 cm]{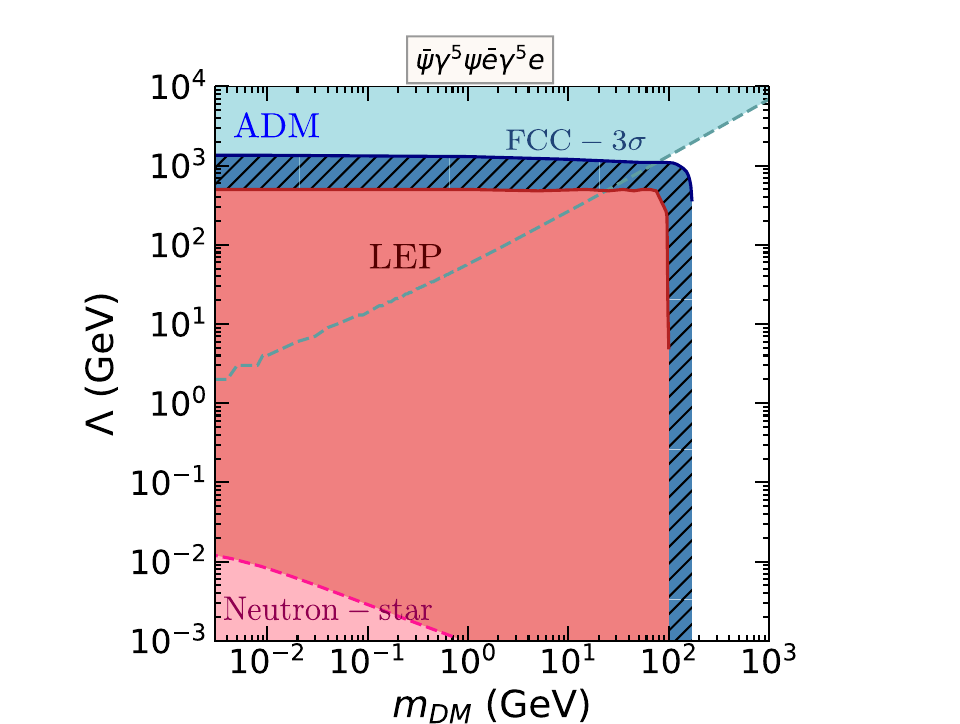}
	\end{subfigure}
	\caption{Same as Fig.~\ref{fig:exc1} but for {pseudo-scalar$-$scalar ($\mathcal{O}_{ps}\equiv$ $ \frac{1}{\Lambda^2}\bar{\psi}i\gamma^5\psi\bar{e}e$ ) (left) and pseudo-scalar$-$pseudo-scalar ($\mathcal{O}_{pp}\equiv$$\frac{1}{\Lambda^2}\bar{\psi}\gamma^5\psi\bar{e}\gamma^5e$) (right)} type interaction of DM with electrons.}
	\label{fig:exc5}
\end{figure}

 We present the overall picture of the allowed and excluded parameter spaces for DM-electron interaction in Figs.~\ref{fig:exc1}\:-\ref{fig:exc5}. Each of the shaded regions within the presented figures signify exclusion due to specific hypotheses. For instance, the blue shaded region eliminates the possibility of dark matter being asymmetric, while the dark red portion is excluded based on the outcomes of the LEP mono-photon searches, and so on. Interestingly, for $\mathcal{O}_{ss},\mathcal{O}_{sp}, \mathcal{O}_{aa}, \mathcal{O}_{av}$, where $\chi\bar{\chi}\to e^+e^-$ occurs through velocity suppressed $p$-wave ($s$-wave annihilation is either absent or has $m_q^2$ suppression), the ADM requirement excludes a larger parameter space in the $\Lambda-m_{\chi}$ plane. It is important to note here that the other exclusions, which are not restricted to the ADM hypothesis, hold for any DM with these effective interactions, and all the regions above the shaded region (except blue) are allowed. Our study reveals that, though the constraints from the NS or WDs on the DM-lepton interactions are overall strong, the bounds coming from the mono-photon searches at the LEP experiments surpasses all those in the sensitive region, $m_\chi\lesssim 100$ GeV. 
 Considering the ADM hypothesis, we find that most of the operators fail to provide an ADM solution for any value of $\Lambda$ in the region $m_\chi\lesssim 100$ GeV, sensitive to the LEP experiment. Only for operators $\mathcal{O}_{vv}$, $\mathcal{O}_{av}$, and $\mathcal{O}_{ps}$, the bounds are somewhat weak $\sim$ 50 GeV, while for $\mathcal{O}_{va}$, $\mathcal{O}_{pt}$, $\mathcal{O}_{sp}$, and $\mathcal{O}_{pp}$ they are much weaker $\sim$10 GeV. Notice that, for the interactions $\mathcal{O}_{ss}$, $\mathcal{O}_{sp}$, $\mathcal{O}_{ps}$, and $\mathcal{O}_{pp}$ the LEP bounds are the only ones that can completely exclude ADM possibility of a certain mass.

The blue striped region in Figs.~\ref{fig:exc1}\:-\ref{fig:exc5} presents the 3-$\sigma$ reach of the upcoming FCC-ee experiment in the mono-photon search channel in the $m_\chi-\Lambda$ plane. The pattern of the probing parameter region is similar to the LEP mono-photon bounds, as expected. However, owing to higher COM energy, the reach extends to $m_\chi\sim$ 175 GeV and constraints may strengthen by a factor of 2 or so.


\newpage
\section{Summary and Outlook}
\input{summary.tex}

\section*{Acknowledgements}
The authors are thankful to Dr. Nishita Desai for useful discussions and suggestions. The authors acknowledge support from the Department of Atomic Energy (DAE), Government of India, under Project Identification No. RTI4002. BD acknowledges partial support through a Swarnajayanti Fellowship of the Department of Science and Technology (DST), Government of India. AR thanks the support of Monash University, where he moved at the end of the writing stage of this manuscript, for support towards completing this work.

\newpage
	\appendix
	\section{Expressions of $\langle\sigma v\rangle$ for different operators}
	The expressions of the annihilation rates for the operators in Table \ref{tab:operators} can be obtained as,
	\br
	\sigma^{\mathcal{O}_{ss}}v&=&\frac{3m_{\chi}^2}{8\pi\Lambda^4}\sum_{q}\left(1-\frac{m_q^2}{m_{\chi}^2}\right)^{3/2}v^2\\
	\sigma^{\mathcal{O}_{pp}}v&=&\frac{3m_{\chi}^2}{\pi\Lambda^4}\sum_{q}\left(1-\frac{m_q^2}{m_{\chi}^2}\right)^{1/2} \left[1+\frac{v^2}{8}\left(\frac{2m_{\chi}^2-m_q^2}{m_{\chi}^2-m_q^2}\right)\right]\\
	\sigma^{\mathcal{O}_{sp}}v&=&\frac{3m_{\chi}^2}{4\pi\Lambda^4}\sum_{q}\left(1-\frac{m_q^2}{m_{\chi}^2}\right)^{1/2}v^2\\
	\sigma^{\mathcal{O}_{ps}}v&=&\frac{3m_{\chi}^2}{4\pi\Lambda^4}\sum_{q}\left(1-\frac{m_q^2}{m_{\chi}^2}\right)^{3/2}\left[1+v^2\left(\frac{2m_q^2-m_{\chi}^2}{m_{\chi}^2-m_q^2}\right)\right]\\	
	\sigma^{\mathcal{O}_{vv}}v&=&\frac{3m_{\chi}^2}{2\pi\Lambda^4}\sum_{q}\left(1-\frac{m_q^2}{m_{\chi}^2}\right)^{1/2}\left[\left(2+\frac{m_q^2}{m_{\chi}^2}\right)
	+v^2\left(\frac{8m_{\chi}^4-4m_q^2m_{\chi}^2+5m_q^4}{24m_{\chi}^2(m_{\chi}^2-m_q^2)}\right)\right]\\
	\sigma^{\mathcal{O}_{aa}}v&=&\frac{3m_{\chi}^2}{2\pi\Lambda^4}\sum_{q}\left(1-\frac{m_q^2}{m_{\chi}^2}\right)^{1/2}\left[\frac{m_q^2}{m_{\chi}^2}
	+v^2\left(\frac{8m_{\chi}^4-22m_q^2 m_{\chi}^2+17m_q^4}{24m_{\chi}^2(m_{\chi}^2-m_q^2)}\right)\right]\\
	\sigma^{\mathcal{O}_{va}}v&=&\frac{3m_{\chi}^2}{\pi\Lambda^4}\sum_{q}\left(1-\frac{m_q^2}{m_{\chi}^2}\right)^{1/2}\left[\left(1-\frac{m_q^2}{m_{\chi}^2}\right)+\frac{v^2}{24}\left(4+5\frac{m_q^2}{m_{\chi}^2}\right)\right]\\
	\sigma^{\mathcal{O}_{av}}v&=&\frac{m_{\chi}^2}{4\pi\Lambda^4}\sum_{q}\left(1-\frac{m_q^2}{m_{\chi}^2}\right)^{1/2}\left(2+\frac{m_q^2}{m_{\chi}^2}\right)v^2\\
	\sigma^{\mathcal{O}_{tt}}v&=&\frac{3m_{\chi}^2}{\pi\Lambda^4}\sum_{q}\left(1-\frac{m_q^2}{m_{\chi}^2}\right)^{1/2}\left[\left(1+2\frac{m_q^2}{m_{\chi}^2}\right)
	+v^2\left(\frac{4m_{\chi}^4-11m_q^2 m_{\chi}^2+16m_q^4}{24m_{\chi}^2(m_{\chi}^2-m_q^2)}\right)\right]\\
	\sigma^{\mathcal{O}_{pt}}v&=&\frac{6m_{\chi}^2}{\pi\Lambda^4}\sum_{q}\left(1-\frac{m_q^2}{m_{\chi}^2}\right)^{1/2}\left[\left(1-\frac{m_q^2}{m_{\chi}^2}\right)+\frac{v^2}{24}\left(4+11\frac{m_q^2}{m_{\chi}^2}\right)\right]	
	\er
	
	\bibliographystyle{JHEP}
	\bibliography{adm.bib}
\end{document}

%% file: summary.tex
In this paper, we present a detailed study of the current status of the effective interactions of ADM with quarks and leptons, with a future projection of the sensitivity of FCC-ee in probing leptophilic ADM. We start with the ADM-quark interactions, where the DD constraints often become most stringent. Considering the strongest exclusions from different DD experiments, we find the excluded and allowed region corresponding to each interaction in the $\Lambda-m_\chi$ plane. However, for several effective interactions, the DM-nucleon scattering cross-sections are suppressed. In these cases, the constraints from monojet searches are dominant. We calculate the exclusions from the monojet searches using the most recent LHC measurements and including detector effects. We find that, for scalar ($\mathcal{O}_{ss}\equiv \frac{1}{\Lambda^2}\bar{\psi}\psi\bar{e}e$) and vector ($\mathcal{O}_{vv}\equiv\frac{1}{\Lambda^2}\bar{\psi}\gamma^{\mu}\psi\bar{e}\gamma_{\mu}e$) type of interactions, the possibility of ADM is ruled out almost up to $\sim$ 1 TeV. Whereas, for other types of interactions, ADM of mass larger than a few hundred GeV are still allowed in a narrow range of $\Lambda$. None of the searches exclude the viability of ADM above $\sim$ 1 TeV. Note that the exclusion regions from the experimental searches are not restricted to the ADM but are also applicable to any DM with the interactions in Table~\ref{tab:operators}.

Next, we concentrate on the ADM-lepton interactions. In this case, the strongest bounds generally come from NS or WD studies. However, we found that the constraints from the monophoton searches at the LEP experiment, which we have calculated in detail for each interaction, overshoot those from the NS or WD for all types of interactions in the $\lesssim$100 GeV mass range. According to our study, the possibility of a leptophilic ADM is ruled out by the LEP experiment up to 100 GeV mass range of the ADM, except for the vector$-$axial-vector operator ($\mathcal{O}_{va}\equiv$ $\frac{1}{\Lambda^2}\bar{\psi}\gamma^{\mu}\psi\bar{e}\gamma_{\mu}\gamma^5e$), the pseudo-tensor$-$tensor operator ($\mathcal{O}_{pt}\equiv$$\frac{1}{\Lambda^2}\bar{\psi}\sigma^{\mu\nu}i\gamma^5\psi\bar{e}\sigma_{\mu\nu}e$), the scalar$-$pseudo-scalar operator ($\mathcal{O}_{sp}\equiv \frac{1}{\Lambda^2}\bar{\psi}\psi\bar{e}i\gamma^5e$), and the pseudo-scalar$-$pseudo-scalar operator ($\mathcal{O}_{pp}\equiv\frac{1}{\Lambda^2}\bar{\psi}\gamma^5\psi\bar{e}\gamma^5e$) mediated interactions, where the constraints are loose $\sim$ $\mathcal{O}$(10) GeV. Again, the constraints apply to any DM having the considered effective interactions (Table~\ref{tab:operators}).

As a possible successor of $e^+e^-$ colliders, the FCC-ee experiment will be interesting to probe the still viable regions of the ADM-lepton interactions. In our analysis, we calculate the region of parameter spaces where the monophoton searches at the FCC-ee will obtain the 3-$\sigma$ sensitivity. This study indicates that FCC-ee can probe up to $\sim$ 200 GeV mass range of any general kind of DM for EFT expansion scale up to $\sim$1 TeV, which, though not a significant improvement, seems attractive from the aspect of possible FCC-ee physics goals.

It is interesting to observe that collider searches have effectively 
	ruled out a significant portion of the parameter space for DM effective 
	interactions, while some leptophilic scenarios still remain viable and 
	can be tested in the future FCC-ee. However, the exclusion of a region in 
	the EFT framework in terms of $m_{\chi}$-$\Lambda$ does not imply that 
	DM particles of mass $m_{\chi}$ are no longer viable. This exclusion 
	essentially means that the interactions of the SM and DM via the heavy 
	mediators (integrated out in the EFT prescription) are excluded. 
	However, the viability of light mediators remains still valid, which may 
	bring non-trivial changes in the results regarding the existence 
	of lower mass DM. It definitely demands a dedicated study.